\documentclass[twocolumn]{aastex6}
\submitted{Submitted to AJ on December 21, 2018, last revised \today}
\usepackage{graphicx}
\usepackage{latexsym}
\usepackage{amssymb}
\usepackage{amsmath}

\usepackage{url}
\usepackage{color, colortbl}

\newcommand{\mj}{$M_{\textrm{\scriptsize Jup}}$~}

\definecolor{aqua}{rgb}{0.0,0.67,1.0}
\definecolor{Gray}{gray}{0.75}
\definecolor{Black}{gray}{1.0}
\hypersetup{citecolor=aqua, linkcolor=aqua}

\begin{document}

\title{Discovery of Six Optical Phase Curves with \textit{K2}}
\author{Prajwal Niraula\altaffilmark{1,2}, Seth Redfield\altaffilmark{2}, Julien de Wit\altaffilmark{1}, Fei Dai\altaffilmark{3,4}, Ismael Mireles\altaffilmark{2}, Dilovan Serindag\altaffilmark{5}, Avi Shporer\altaffilmark{3}}

\altaffiltext{1}{Department of Earth, Atmospheric and Planetary Sciences, MIT, 77 Massachusetts Avenue, Cambridge, MA 02139, USA; Corresponding author: pniraula@mit.edu}
\altaffiltext{2}{Astronomy Department and Van Vleck Observatory, Wesleyan University, Middletown, CT 06459, USA}
\altaffiltext{3}{Department of Physics and Kavli Institute for Astrophysics and Space Research, Massachusetts Institute of Technology, Cambridge, MA 02139,
USA }
\altaffiltext{4}{Department of Astrophysical Sciences, Peyton Hall, 4 Ivy Lane,
Princeton, NJ 08540 USA}
\altaffiltext{5}{Leiden Observatory, Leiden University, P.O. Box 9513, 2300 RA Leiden, The Netherlands}
\altaffiltext{}{Submitted\today, Accept \today, printed \today}

\begin{abstract} 
We have systematically searched for the phase curves among the planets discovered by \textit{K2}. Using the  reported planetary parameters, we  screen out the best potential candidates, and examine their light curves in detail. For our work, we consider light curves from two different detrending pipelines - \texttt{EVEREST} and \texttt{K2SFF}. In order to remove stellar variability and systematics, we test three different filtering techniques: spline, phasma (median-filtering) and Butterworth (harmonics filtering), and use Butterworth filtered light curves for the subsequent analysis. We have identified 6 previously unreported phase curves among the planets observed with \textit{K2}: K2-31b, HATS-9b, HATS-11b, K2-107b, K2-131b, and K2-106b. The first four of these are hot Jupiters for which we find the photometric masses consistent with their RV-based masses within 2$\sigma$, 1$\sigma$, 1$\sigma$, and 3$\sigma$ respectively with comparatively low geometric albedos, while the last two are ultra-short period super-Earths with phase curves dominated by reflective and thermal components. We also detect a secondary eclipse in HATS-11b at 62$\pm$12 ppm. We thus deem it to be possible to validate the planetary nature  of selected \textit{K2}, and suggest similar vetting could be used for the ongoing \textit{TESS} mission. 
\end{abstract}


\section{INTRODUCTION} 
\label{sec:intro}
Phase curve analysis has established itself as an important tool in characterizing exoplanets. It is commonly used to study close-in massive planets, opening up a window to independently characterize the planetary parameters such as eccentricity, geometric albedo, longitudinal temperature distribution, cloud coverage as well as planetary mass (for a detail review, see \citet{shporer2017}, or \citet{parmentier2017}). While infrared windows are often ideal for observing the thermal phase curve of planets as planetary emission often peak in such bandpass \citep[e.g.][]{harrington2006, knutson2007, adams2018}, the limited number of space based infrared facilities has been the primary bottleneck for such studies. This is where space-based photometric missions such as Convection, Rotation, and planetary Transits \citep[\textit{CoRoT};][]{baglin2002}, and \textit{Kepler} \citep{borucki2010} have come to play a crucial role through optical phase curves. While leading in the discoveries of the exoplanets, these mission have also significantly increased the number of phase curve detections. For instance with \textit{Kepler},  photometric time series were obtained for thousands of targets with unprecedented precision. For reference, a combined differential photometric precision of $29.0$  ppm was reported for Kp = 11.5 to 12 mag stars  in the long cadence mode \citep{gilliland2011}. This has already led to the discovery of robust phase curves for more than 20 transiting \textit{Kepler} planets \citep{borucki2009, welsh2010, batalha2011, barclay2012, esteves2015, angerhausen2015}. Additionally, the precise photometry has been used to estimate RV-independent masses \citep{faigler2013, esteves2013}, validating the planetary nature of transiting objects \citep{quintana2013, ojeda2013} well as for the discovery non-transiting systems \citep{faigler2012, millholland2017}. 

Despite operating only on two reaction wheels, the revamped \textit{K2} mission is able to achieve photometric precision on par with  the original \textit{Kepler} mission. This has been possible not only due to ingenious mission redesign \citep{howell2014} but also because of a host of tools that have been developed in response to the unique data challenges caused primarily by, but not limited to, the telescope drift \citep{vanderburg2014, luger2016}. Despite \textit{K2}'s relatively short baseline of $\sim$80 days (compared to the primary mission's four year baseline), its high precision observation of numerous bright targets (V$\sim$10 mag) translates into good opportunities for detecting phase curves.

Yet, there are unique challenges in studying phase curves with a `short' observation baseline. Not only does the shorter length of observation make it particularly difficult when it comes to disentangling the phase curve from quasi-periodic signals such as those arising from spot modulation, but non-periodic effects such as thermal settling or edge effects will disproportionately distort the final obtained signal. In the case of \textit{K2}, the required aggressive post processing of the data to correct the systematics can also affect the astrophysical signal under consideration. Not to mention, there are gaps in our understanding of the physics behind phase curves. Intriguing questions surrounding the existence and the cause of the third harmonics observed in systems such as HAT-P-7b and Kepler-13Ab still eludes a clear explanation \citep{shporer2014, esteves2015, cowan2017}.  A handful of the optical phase curves have been observed with significant asymmetries, which have been attributed to scattering due to inhomogeneous clouds  \citep{demory2013, shporer2015}. Meanwhile, \citet{armstrong2016} measured the temporal variations of the optical phase curve of HAT-P-7b, which comes with the prescription for the climatic variability in planets, thereby undermining the classical picture of a consistent signal present throughout the time series. Similarly, given the small signal amplitude, correlated noise as well as dilution can dramatically affect the inferred conclusions we derive from the phase curves  \citep[see discussions surrounding Kepler-91b in ][]{esteves2013,lillobox2014}. This all points to the complex world of planetary atmospheres hidden under the simple phase curve models, robust characterization of which would require more precise data.

Despite these challenges, there are already three reported planetary phase curves in \textit{K2} data: Qatar~2b \citep{dai2017b}, K2-141b \citep{malavolta2018}, and WASP-104b \citep{mocnik2018}. Hot Jupiters like Qatar-2b and WASP-104b have light curves exhibiting ellipsoidal variation and Doppler boosting consistent with their radial velocity (RV) based masses. On the other hand,  K2-141b, an ultra short period super-Earth, has a measurable phase curve dominated by reflective and thermal components. As more than 350 planets have been discovered by \textit{K2}, we have undertaken a project to systematically search for phase curves among the selected \textit{K2} light curves where phase curve detection is feasible.

In  $\S$2 of this work, we introduce the process used to  screen out the targets for the phase curves among the \textit{K2} planets. In $\S$3, we present the pipeline used to process the data including discussion on different filtering processes, and in $\S$4 we perform Signal Injection and Retrieval test to probe the strengths and weaknesses of our pipelines. Under $\S$5, we discuss the model framework used for the phase curve, and the fitting procedures whereas the results are presented in $\S$6. This is followed by the reportings on secondary eclipse in $\S$7. In $\S$8 and $\S$9, we present the scientific insights gained through our work, which is followed by the conclusion. We present the the light curves used in our data analysis in the Appendix.

\section{Potential Candidates}
For this study, we have only considered the confirmed exoplanets. Our search included 382 exoplanets which were observed by \textit{K2}, all of which were catalogued in NASA Exoplanet Archive\footnote{\href{https://exoplanetarchive.ipac.caltech.edu}{https://exoplanetarchive.ipac.caltech.edu}} as of December 18, 2018 with the \textit{K2} flag. While the \textit{K2} mission itself has come to an end, the data from the mission is still being processed and discovery numbers are expected to increase as candidates continue to be validated with follow-up observations. Phase curve studies, such as the analysis presented in this paper, can also be applied to other photometric exoplanet surveys such as Transiting Exoplanet Search Survey \citep[\textit{TESS;}][]{ricker2015}, which is expected to find as many as $\sim$10$^4$ planets primarily in short period orbits \citep{huang2018}.

As the possibility of the detection of phase curves primarily boils down to the precision of the light curve, we filter out  the suitable candidates by estimating the magnitude of the combined signal against the obtainable precision of the light curve. We use the parameters recorded in the database to estimate the equivalent amplitude  of the phase curve signal. When some of the values  were missing, estimates were made based on other available parameters of the planet. Using the combined amplitude of all four different effects i.e. reflective, thermal, ellipsoidal and Doppler components, we calculate the expected signal to noise ratio using an estimator for the precision of the light curve as below:

\begin{equation}
\begin{aligned}
\label{eqn:CalcAmp}
\mathrm{SNR} &= \frac{{((A_{Re\!f}+A_{Th})^2+A_{Ell}^2+A_{Dop}^2})^{\frac{1}{2}}\cdot N^{\frac{1}{2}}}{\sqrt{2}\sigma} \\
 &= \frac{ A_{Eqv} \cdot N^{\frac{1}{2}}}{\sqrt{2} \sigma} 
\end{aligned}
\end{equation}
where $N$ is the number of observed photometric points set to 3500 (roughly 30 minutes bin over 80 days observation period), and $\sigma$ is the precision expected to be determined by the brightness of the target in \textit{Kepler} bandpass using pre-flight estimates. The equivalent amplitude is considered to be the amplitude of the sinusoidal signal which has power equivalent to the combined elements of the phase curve:
\begin{equation}
\begin{aligned}
\label{eqn:CalcAmp}
A_{Eqv} &= ((A_{Re\!f}+A_{Th})^2 + A_{Ell}^2 + A_{Dop}^2)^{\frac{1}{2}} 
\end{aligned}
\end{equation}

To estimate the reflective component ($A_{Re\!f}$), we assume the geometric albedo ($A_g$) of 0.4 and evaluate the reflective component ($A_{Re\!f}$) as follows:
\begin{equation}
A_{Re\!f} = A_g \left(\frac{R_p/R_*}{a/R_*}\right)^2,
\end{equation}
where $R_p/R_*$ is the scaled radius of the planet, and $a/R_*$ is the scaled semi-major axis. Similarly, the thermal variation ($A_{Th}$) is calculated as: 
\begin{equation}
A_{Th} =  \left(\frac{R_p}{R_*}\right)^2 \frac{\int B(T_{Day}) R(\lambda) d \lambda}{\int B(T_{*}) R(\lambda)d \lambda},
\end{equation}
where $B(T)$ is the Planck's black body radiation law corresponding to temperature $T$,  $R(\lambda$) is the response function of \textit{Kepler/K2}, and $T_{Day}$ is the day-side temperature of the planet, which is estimated as following:
\begin{equation}
T_{Day} = T_{e\!f\!f}\sqrt{\frac{1}{a/R_*}}\left[f(1-A_B\right)]^{\frac{1}{4}} ,
\label{eqn:lopez2007}
\end{equation}
where $T_{e\!f\!f}$ is the effective temperature of the host star, $A_B$ is the Bond albedo set at 0.6 following a Lambertian sphere relation ($A_B$ = $\frac{3}{2} A_g$), and $f$ is a proxy variable for re-circulation set at 2/3 corresponding to the case where only the day-side is re-radiating  \citep{lopez2007}.

\begin{figure*}[ht]
\includegraphics[width=0.97\textwidth]{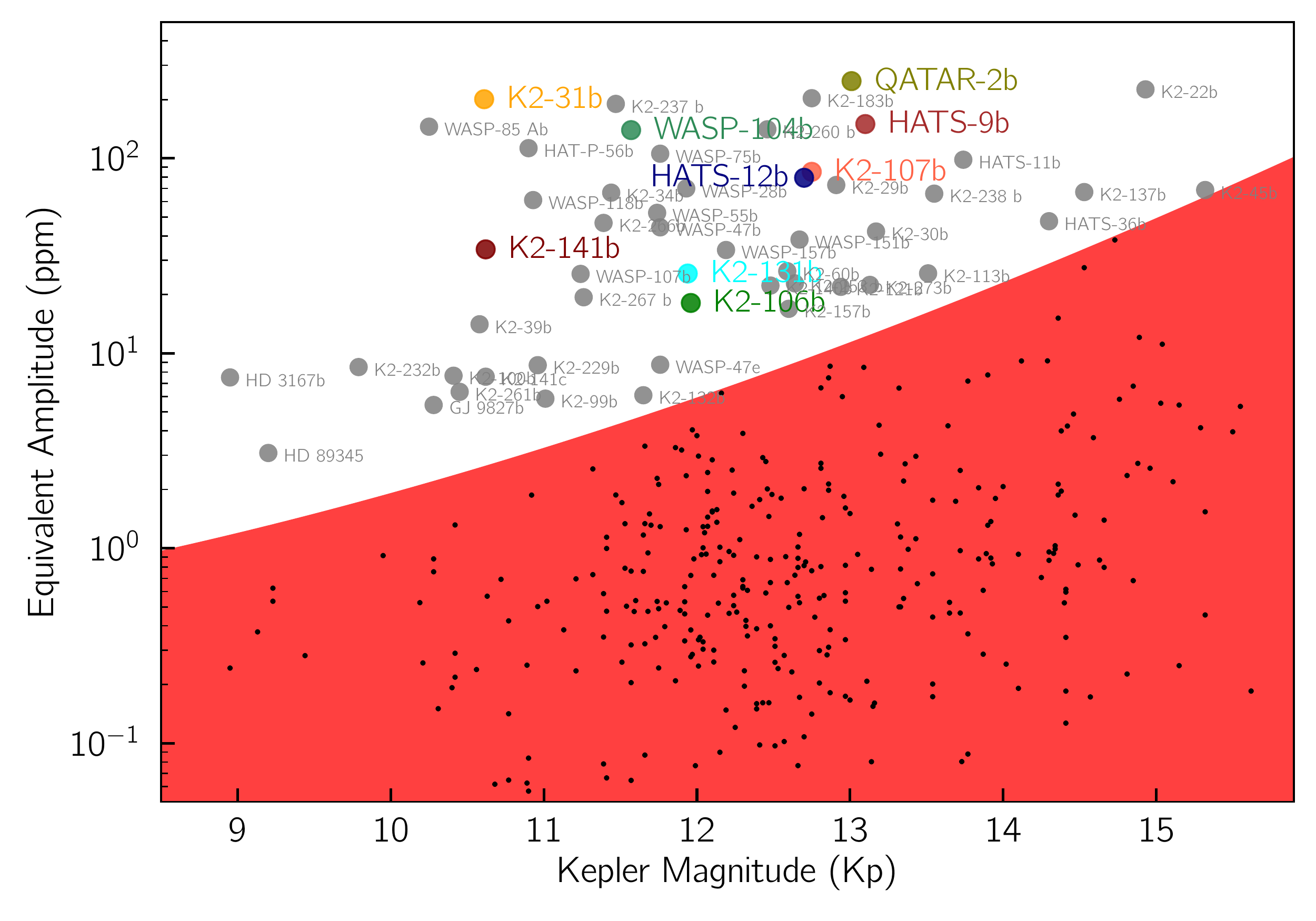}
\caption{\label{fig:best_candidates}  Expected signal amplitude ($A_{eqv}$) versus the \textit{Kepler} magnitude for the \textit{K2} detected planets. The red region contains the planet for which SNR detection is expected to be less than 3$\sigma$. Out of 382 planets, 52 could have detectable phase signal above 3$\sigma$ level and are listed in \autoref{table:best_candidates}. Among these,  9 planets have detected phase curves, all of which are drawn in colors other than grey to improve the readability of the graph.} 
\end{figure*}

\begin{deluxetable*}{lcccccccccccc}
\tablecaption{\label{table:best_candidates} All \textit{K2} targets analyzed for the phase curves with various relevant parameters.}
\tablehead{
\colhead{\textbf{Identifier}}&
\colhead{\textbf{EPIC ID}}&
\colhead{\textbf{Kp} (Mag)} &
\colhead{\textbf{Campaign}} &
\colhead{ \textbf{Period} (Days)}&
\colhead{\textbf{$Rp/R_{*}$}}&
\colhead{ \textbf{$A_{Re\!f}$}}& 
\colhead{\textbf{$A_{Th}$}}& 
\colhead{\textbf{$A_{Ell}$}} & 
\colhead{\textbf{$A_{Dop}$}} & 
\colhead{\textbf{$A_{Eqv}$}} & 
\colhead{\textbf{SNR}} 
}
\startdata
\textbf{K2-31b} & 204129699 & 10.6 & 2 & 1.258 & 0.135 & 199.2 & 1.4 & 11.9 & 4.7 & 201 & 228.9\\
\rowcolor{Gray}
 WASP-85 Ab & 201862715 & 10.3 & 1 & 2.656  & 0.163 & 144.5 & 0.80 & 2.2 & 2.1 & 145 & 200.2\\
 K2-237b & 229426032 & 11.5 & 11& 2.181 & 0.118 & 184.1  & 5.8 & 8.8 & 2.4 & 190 & 132.9\\
\rowcolor{Gray}
HAT-P-56b & 202126852 &  10.9 & 0 & 2.791 & 0.105 & 109.5 & 3.0 & 7.0 & 2.8 &113 &109.3 \\
\textbf{WASP-104b$^{a}$} & 248662696 & 11.6 & 14 & 1.755 &  0.121 & 138.4 & 0.9 & 5.8 & 2.7 &140 & 91.9\\
\rowcolor{Gray}
\textbf{QATAR-2b$^{b}$} & 21275629 & 13.0 & 6 & 1.337 &  0.162 & 247.6 & 0.4 & 18.5 & 8.3 & 249 &65.1\\
K2-183b & 211562654 & 12.8 & 5 & 0.469 & 0.027 & 152.8 & 50.6 & 1.9 & 0.0 & 203 & 63.2\\
\rowcolor{Gray}
WASP-75b & 206154641 & 11.8 & 3& 2.484 & 0.103 &103.8 & 1.7 & 4.2 & 1.7 & 106 & 61.8 \\
WASP-118b & 220303276 & 10.9 & 8 &4.046 &  0.082 & 59.9 & 1.2 & 1.5 & 0.6 & 61 & 58.1\\
\rowcolor{Gray}
K2-260 b & 246911830 &  12.5 & 13&2.627 &  0.097 & 135.3 & 5.3 & 7.9 & 1.9 & 141 & 52.9\\
K2-34b & 212110888 &  11.4 & 5, 16 &2.996 &  0.088 & 65.5 & 0.9 & 6.3 & 2.5 & 67 &47.3\\
\rowcolor{Gray}
\textbf{K2-141b$^{c}$} & 246393474 &  10.6 & 12 &0.280 &0.020 & 31.6  & 2.4 & 2.8 & 0.1 & 34 &38.6 \\
\textbf{HATS-9b} & 217671466 & 13.1 & 7 &1.915 &  0.083 & 145.5 & 3.7 & 13.6 & 1.8 &150 & 37.0\\
\rowcolor{Gray}
WASP-28b & 246375295 & 11.9 & 12 & 3.407 &  0.116 & 69.7 & 0.4 & 1.5 & 1.4 &70 & 36.9\\
K2-266b & 248435473 & 11.4 & 14 & 0.659 &  0.043 & 46.2 & 0.4 & 1.1 & 0.2 &47 & 34.2\\
\rowcolor{Gray}
WASP-55b & 212300977 & 11.7 &  6 & 4.466 &  0.125 & 52.4 & 0.1 & 0.5 & 0.8 & 53 &32.2\\
\textbf{K2-107b} & 216468514 &  12.8 & 7 &3.314 &  0.083 & 83.3 & 1.8 & 4.1 & 1.1 & 85 &26.6\\
\rowcolor{Gray}
WASP-47b & 206103150 & 11.8 & 3 & 4.161 & 0.102 & 44.1 & 0.1 & 1.6 & 1.8 & 44 & 25.9\\
\textbf{HATS-12b} & 218131080 &  12.7 & 7 & 3.143 & 0.063 & 73.2 & 4.2 & 17.8 & 2.8 & 79 &25.6\\
\rowcolor{Gray}
WASP-107b & 228724232 & 11.2 & 10 &5.721 & 0.145 & 25.5 & 0.0 & 0.0 & 0.3 & 26 & 20.4\\
K2-29b & 211089792 & 12.9 & 4 & 3.259 &  0.142 & 72.9 & 0.7 & 0.9 & 1.4 & 73 &20.4\\
\rowcolor{Gray}
HD 3167b & 220383386 & 9.0 & 8 & 0.957 &  0.017 & 7.3 & 1.4 & 0.4 & 0.0 & 7.5 &19.3\\
K2-39b & 206247743 &  10.6 & 3 & 4.605 &  0.019 & 12.7 & 0.4 & 5.0 & 0.2 & 14 &16.4\\ 
\rowcolor{Gray}
HATS-11b & 216414930 & 13.7 & 7 & 3.619 & 0.107 & 97.1 & 1.2 & 3.1 & 1.3 & 98 &15.5\\
K2-267b & 246851721 & 11.3 & 13 & 6.180 & 0.0681 & 19.3 & 0.1 & 0.1 & 0.1 & 19 &15.3 \\
\rowcolor{Gray}
WASP-157b & 212697709 &  12.2 & 6 & 3.952 &  0.094 & 33.8 & 0.1 & 0.5 & 0.8 & 34 &15.2\\
K2-232b & 247098361 & 9.79 & 13 & 11.168 & 0.020 & 8.5 & 0.0 & 0.1 & 0.4 & 8.5 &14.8\\ 
\rowcolor{Gray}
K2-22b & 201637175 & 14.9 & 1 & 0.381 & 0.075 &  205.8 & 1.8 & 87.7 & 9.0 & 226 & 14.5\\
\textbf{K2-131b} & 228732031 & 11.9 & 10 & 0.369 & 0.020 & 23.5 & 2.1 & 1.8 & 0.1 & 26 &13.5\\
\rowcolor{Gray}
WASP-151b & 246441449 & 12.7 & 12 & 4.533  & 0.101 & 38.2 & 0.1 & 0.3 & 0.4 & 38 &12.6 \\
K2-238 b & 246067459 & 13.6 & 12 & 3.205  & 0.080 & 65.1 & 0.6 & 3.8 & 1.3 & 66 &11.8 \\
\rowcolor{Gray}
K2-30b & 210957318 & 13.2 & 4 & 4.100 &  0.127 & 42.1 & 0.0 & 0.5 & 1.0 & 42 &9.9 \\
K2-100b & 211990866 & 10.4 & 5 & 1.674 &  0.027 & 7.5 & 0.1 & 0.1 & 0.0 & 7.7 &9.7 \\
\rowcolor{Gray}
\textbf{K2-106b} & 220674823 & 12.0 & 8 & 0.571 &  0.017 & 16.3 & 1.7 & 2.0 & 0.1 & 18 &9.4 \\
K2-60b & 206038483 & 12.6 & 3 & 3.003 & 0.063 & 26.2 & 0.1 & 1.2 & 0.8 & 26 & 9.1 \\
\rowcolor{Gray}
K2-141c & 246393474 & 10.6 & 12 & 7.749 &  0.094 & 7.6 & 0.0 & 0.0 & 0.0 & 7.6 & 8.6 \\
K2-140b & 228735255 & 12.5 & 10 & 6.569 &  0.114 & 22.2 & 0.0 & 0.4 & 1.4 & 22 &8.3 \\
\rowcolor{Gray}
K2-229b & 228801451 & 11.0 & 10 & 0.584 &  0.014 & 8.2 & 0.5 & 0.4 & 0.0 & 8.7 &8.2 \\
K2-261b & 201498078 & 10.5 & 14 & 11.633 & 0.053 & 6.3 & 0.0 & 0.1 & 0.2 & 6.3  & 7.9 \\
\rowcolor{Gray}
K2-253 b & 228809550 & 12.6 & 10 & 4.002 &  0.105 & 22.8 & 0.0 & 0.0 & 0.1 &23 & 7.7 \\
GJ 9827b & 246389858 & 10.3 & 12 & 1.209 &  0.025 & 5.4 & 0.0 & 0.1 & 0.1 & 5.4 & 7.4 \\
\rowcolor{Gray}
HD 89345 & 248777106 & 9.2 & 14 &11.814  &  0.038 & 3.1 & 0.0 & 0.1 & 0.1 & 3.1 & 7.0 \\
K2-121b & 211818569 & 12.9 & 5 & 5.186  &  0.109 & 21.9 & 0.0 & 0.0 & 0.1 & 22 &6.0 \\
\rowcolor{Gray}
K2-137b & 228813918 & 14.5 & 10 & 0.180  &  0.018 & 17.0 & 0.2 & 64.7 & 5.1 & 67 &5.8 \\
K2-157b & 201130233 & 12.6 & 10 & 0.365  &  0.011 & 13.9 & 2.9 & 1.8 & 0.0 & 17 &5.8 \\
\rowcolor{Gray}
K2-273b & 211919004 & 13.1 & 5,16,18 &11.716 &  0.0484 & 22.3 & 0.1 & 0.2 & 0.0 & 22 & 5.4 \\
K2-99b & 212803289 & 11.0 & 6,17 & 18.249 &  0.042 & 5.8 & 0.0 & 0.5 & 0.7 & 5.9 &5.3 \\
\rowcolor{Gray}
WASP-47e & 206103150  & 11.8 & 3 & 0.790 &  0.014 & 8.1 & 0.6 & 1.0 & 0.1 & 8.7 &5.1 \\
HATS-36b & 215969174 & 14.3 & 7 & 4.175 &  0.110 & 47.2 & 0.1 & 3.1 & 4.2 & 48 & 4.9 \\
\rowcolor{Gray}
K2-113b & 220504338 & 13.5 & 8 & 5.818 &  0.091 & 25.6 & 0.0 & 1.1 & 1.8 & 26 &4.7 \\
K2-132b & 228754001 & 11.7 & 10 &9.175 &  0.033 & 5.9 & 0.0 & 1.1 & 0.6 & 6.1 & 3.8 \\
\rowcolor{Gray}
K2-45b & 201345483 & 15.3 & 1 & 1.729 &  0.138 & 69.1 & 0.0 & 0.1 & 0.2 & 69 &3.3
\enddata
\tablenotetext{}{References for optical phase curves in (a) \citet{mocnik2018}, (b) \citet{malavolta2018}, and (c) \citet{dai2017b}}
\tablenotetext{$\dag$}{All the detected phase curves planets are highlighted in bold.}
\end{deluxetable*}

For calculating the amplitude of the ellipsoidal variation $A_{Ell}$, we consider the formulation presented in \citet{morris1985}:
\begin{align}
A_{Ell} &= \alpha_{Ell} \frac{M_p}{M_*} \left(\frac{1}{a/R_*}\right)^3  \sin ^2 i,\\
\alpha_{Ell} &= \frac{0.15 (15+u)(1+g)}{(3-u)},
\end{align}

where $\alpha_{Ell}$ is a constant characterizing tidal distortion, $M_p$ is the mass of the planet, $M_*$ is the mass of the star, $a/R_*$ is the scaled semi-major axis, $i$ is the inclination of the orbit, $u$ is the linear limb-darkening parameter, and $g$ is the gravity-darkening parameter. We determine the value of $u$ and $g$  by linearly interpolating among effective temperature, metallicity and $log~g$ and assuming  turbulence of 2 $\rm{km~s}^{-1}$ from the table provided by \citet{claret2011}.
 
Similarly, the Doppler beaming effect ($A_{Dop}$) is modeled as following:
\begin{align}
&K = \left(\frac{2 \pi G}{P}\right)^{1/3} \frac{M_p \sin i}{M_*^{2/3}\sqrt{1-e^2}},\quad (M_p<<M_*)\\
&A_{Dop} = \alpha_D \frac{K}{c},
\end{align}
where $G$ is the gravitational constant, $P$ is the period, $M_p$ is the mass of the planet, $M_*$ is the mass of the star. $\alpha_D$ is Doppler boosting factor, which is based on the  proposition of \citet{loeb2003}. For this work we use the empirical relation reported by \citet{millholland2017} between the Doppler boosting coefficient ($\alpha_{D}$) and effective stellar temperature ($T_{e\!f\!f}$): 

\begin{equation}
\begin{aligned}
\alpha_{D} = 7.2-(6\times 10^{-4})T_{e\!f\!f}.
\end{aligned}
\end{equation}

\begin{figure}[ht!]
\includegraphics[width=0.49\textwidth]{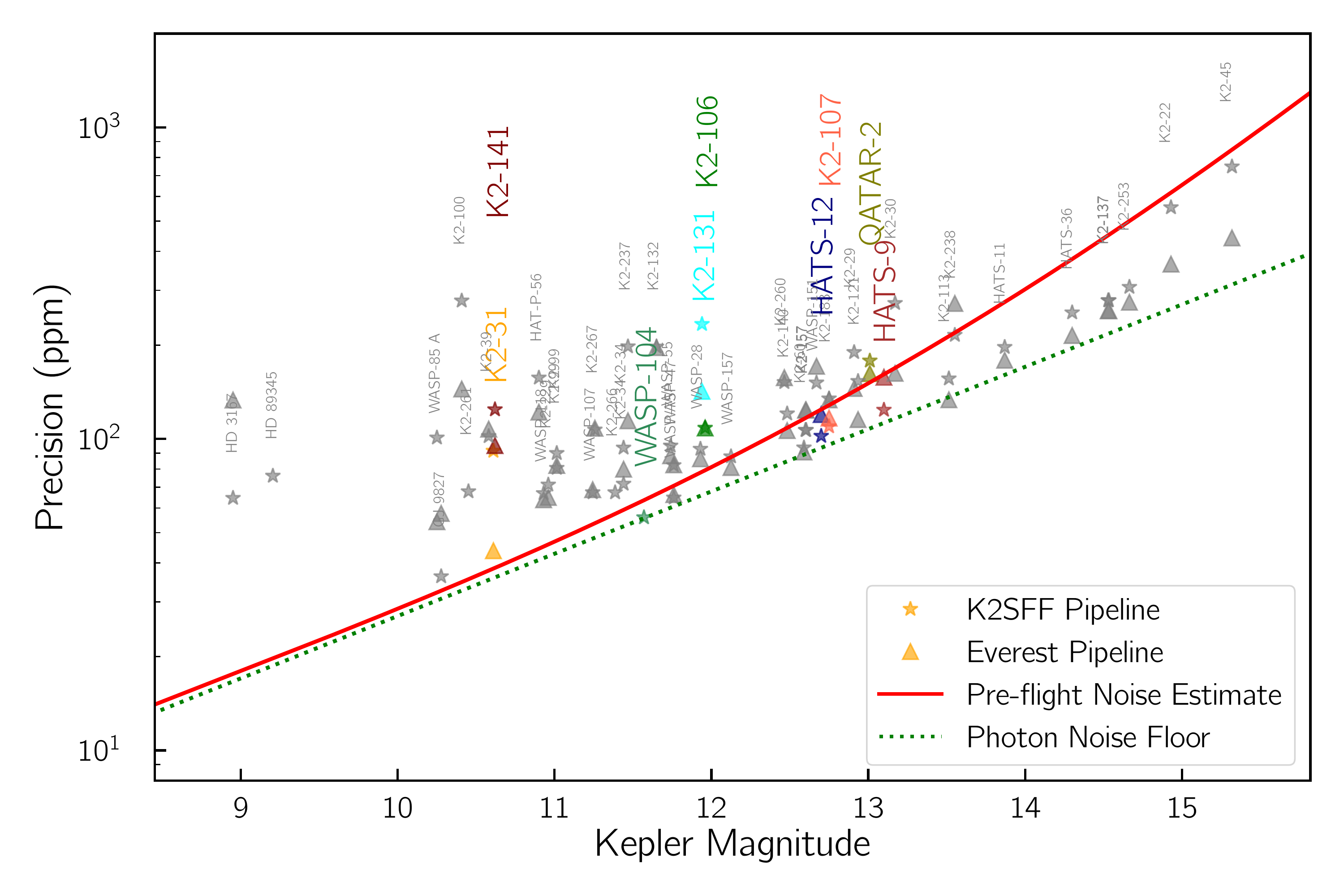}
\caption{\label{fig:ExpectedPrecision} The observed precision for our targets compared to the expected noise level used for SNR estimation. The expected noise level underestimates the scatter for bright targets while generally overestimates the noise for the dimmer targets. }
\end{figure}

\autoref{table:best_candidates} lists 52 \textit{K2} exoplanets arranged in the order of expected signal to noise for phase curve. It also lists the expected individual contributions of all four phase curve components in parts per million (ppm), along other fundamental parameters and information. \autoref{fig:best_candidates} displays predicted phase curve amplitudes ($A_{Eqv}$) as a function of stellar magnitude.  Note that while we expect the ellipsoidal variation and Doppler amplitude to be well constrained around our predicted values, the reflective and thermal components can differ by more than an order of magnitude, depending on the choice of albedo. The choice of high geometric albedo is motivated primarily to cast a wide enough net not to miss any potential candidates. This in turn leads to a high non-detection rate. Additional factors that lead to decreased precision include crowding, imperfect detrending, stellar activity, and other various systematics. Besides, we find that the pre-flight estimates systematically overpredicts the precision for the brighter targets as can be seen in \autoref{fig:ExpectedPrecision}. We trace this back to the use of constant aperture size across all targets.

\section{Data Preparation}
Phase curves signals are often times  weak compared to other astrophysical signals present in the photometric time series. Filtering processes to remove them is therefore a necessary step. By the time the final light curve is obtained, the data goes through multiple processing steps, each of which handles different aspects of the systematics. The official \textit{Kepler} processing tool handles many of the detector and electronic effects \citep{jenkins2010}. However, the pointing induced errors historically have been left up to the exoplanet community to address. 

As a response,  different pipelines were developed by research groups who have diverse research foci. For instance, Kepler Asteroseismic Science Operations Center  \citep[KASOC;][]{handberg2014} pipeline was purposed for astero-seismic related studies, whereas for those interested in the stellar rotational period developed independent pipelines \citep{angus2016}. Similarly, specialized pipelines have been developed to find transits. For our work we consider two different pipelines, \texttt{K2SFF} \citep{vanderburg2014, vanderburg2016} and \texttt{EVEREST} \citep{luger2016, luger2018} primarily due to their ability to produce light curves with high precision. For our work, we have used the scatter of the phase folded light curve as the benchmark for selection criterion for any subsequent analysis. This was motivated by the reasoning that systematics is the single-handedly the most challenging hurdle standing in the way of the phase curve detection.

\subsection{Detrending}
Due to drift of the field in the rolling axis, which was periodically compensated for by thruster firings, pre-detrended light curves from \textit{K2} exhibit a characteristic sawtooth pattern. Pipelines such as \texttt{EVEREST}, and \texttt{K2SFF} have been designed to remove these and other systematics unique to the \textit{K2} data. We particularly focus on using these two pipelines due to their comprehensiveness and the quality of the final products delivered by these pipelines. \texttt{K2SFF} light curves are available for all of our targets, whereas \texttt{EVEREST} is available for all of targets up until Campaign 13 at the time of this publication. All the data is publicly available from the MAST archive.\footnote{\href{https://archive.stsci.edu/k2/data\_search/search.php}{https://archive.stsci.edu/k2/data\_search/search.php}}

\texttt{K2SFF} is a parametric approach to detrending - decorrelating the motion of the centroid with the variation of the magnitude. On the other hand, \texttt{EVEREST} is a Gaussian process based detrending which models the intra-pixel variation to produce the final light curve . Both of the methods produce light curves that are comparable in photometric precision (see \autoref{fig:ExpectedPrecision}). Unfortunately, a comprehensive quantitative comparisons on the performance of these pipelines is beyond the scope of this paper.  Light curves from both pipelines have been used successfully in different analyses. \citet{dai2017b} used the light curve generated by \texttt{EVEREST} pipeline due to its higher precision whereas \citet{malavolta2018} and \citet{mocnik2018} used some variation of \texttt{K2SFF} in producing the final light curves, which were then used for subsequent phase curve analyses. 

The detrending removes most of the power injected at higher frequencies such as those introduced by thruster firing events ($\sim$6 hours), while at  lower frequencies ($>$15 days) other long term systematics still can dominate \citep{vancleve2016, aranzana2018}. Since the phase curve signals lie in the region ($\sim$1 to 5 days) which is usually well separated in the frequency domain from both of these effects, they are minimally distorted, and the treatment by these pipelines are sufficient in most of the cases. However, there are cases where these traditional \textit{K2} pipelines often tend to fail, such as in crowded fields or for bright targets. This in turn has motivated the development of more specialized pipelines to handle crowding typically in a cluster environment \citep{libralato2016} or bright targets that can saturate the pixels \citep{white2017}. Unfortunately, the light curves from these specialized pipelines are not as comprehensively available for most of \textit{K2} targets, therefore we limit our analysis with the light curves available from \texttt{EVEREST} and \texttt{K2SFF}.

In \autoref{fig:ExpectedPrecision}, we compare the observed precision of the light curve against the expected precision. We found that the pre-flight prediction provided for \textit{Kepler} targets\footnote{\href{https://keplergo.arc.nasa.gov/CalibrationSN.shtml}{https://keplergo.arc.nasa.gov/CalibrationSN.shtml}} overestimates the precision for the bright targets, whereas it underestimates the precision for the fainter targets. This error can be traced to the constant read noise error assigned to all of the targets calculated by assuming an aperture size of 20 pixels. For the bright targets the non-linear effects and background contamination poses worse problems than this calculation allows for. While a separate algorithm for detrending bright stars has been explored \citep{white2017}, there might be room for even more optimization. But overall, the assumed precision curve provides a good estimate of the precision expected for our targets.

\begin{figure*}[ht]
\includegraphics[width=0.97\textwidth]{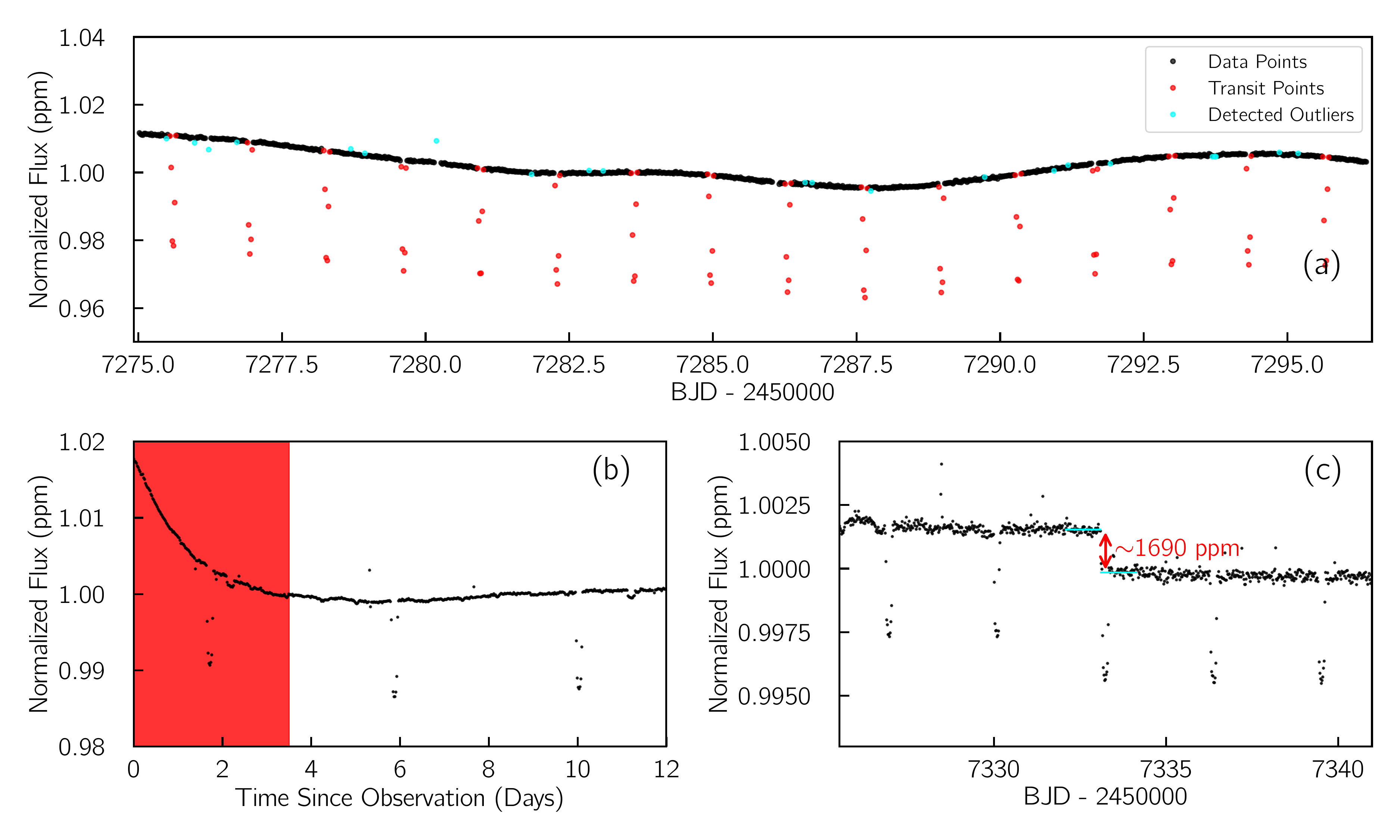}
\caption{\label{fig:anomaly} Figure showing different systematics in the detrended data. (a) Transit points and outliers detected in a portion of QATAR-2 \texttt{EVEREST} detrended light curve.  (b) Potential systematic effects akin to thermal settling observed during the first few days of detrended WASP-47 \texttt{EVEREST} detrended data. (c) An abrupt change is  observed in \texttt{EVEREST} detrended data for HATS-12b possibly due to a change in pixel responsivity. Note that a similar offset is spotted in \texttt{K2SFF} detrended data.}
\end{figure*}

\subsection{Outliers Handling}
The final light curve obtained by the pipelines needs further processing due to the presence of outliers. We initially remove all the data lying outside eight times the inter-quartile (Q3-Q1) range from the median. Following this, we mask out all the transit points found using the transit parameters reported in the NASA exoplanet database. We then locate the outliers through an iterative spline fitting process by excluding data that lie more than 3$\sigma$ away in each iteration. This process is repeated once the light curve is folded, during which outliers occurring during transit events are removed. On average, this led to removal of around 2.5\% of the original data across our targets.

Some of our targets show effects akin to thermal settling at the beginning of the data (see \autoref{fig:anomaly}). However, since these effects are not uniformly present in all our targets, we did not exclude these data points in our analysis.  If multi-campaign observations were available for a target, we combined all of the light curves available to produce the final light curve. \textit{K2} light curves often have a gap typically of a few days at the middle of the observation for data downlinking. These discontinuities are expected and well handled by the detrending pipelines as well as the filtering. For  HATS-12 however, there is an abrupt offset at the middle of the observation (see \autoref{fig:anomaly}) which was observed in both of the detrended light curves. We correct for this apparent offset by modeling the continuum using linear regression at the break point, and applying the relevant offset to generate the final stellar continuum before filtering.

\section{Signal Injection Retrieval Test}
\begin{figure}[ht]
\includegraphics[width=0.48\textwidth]{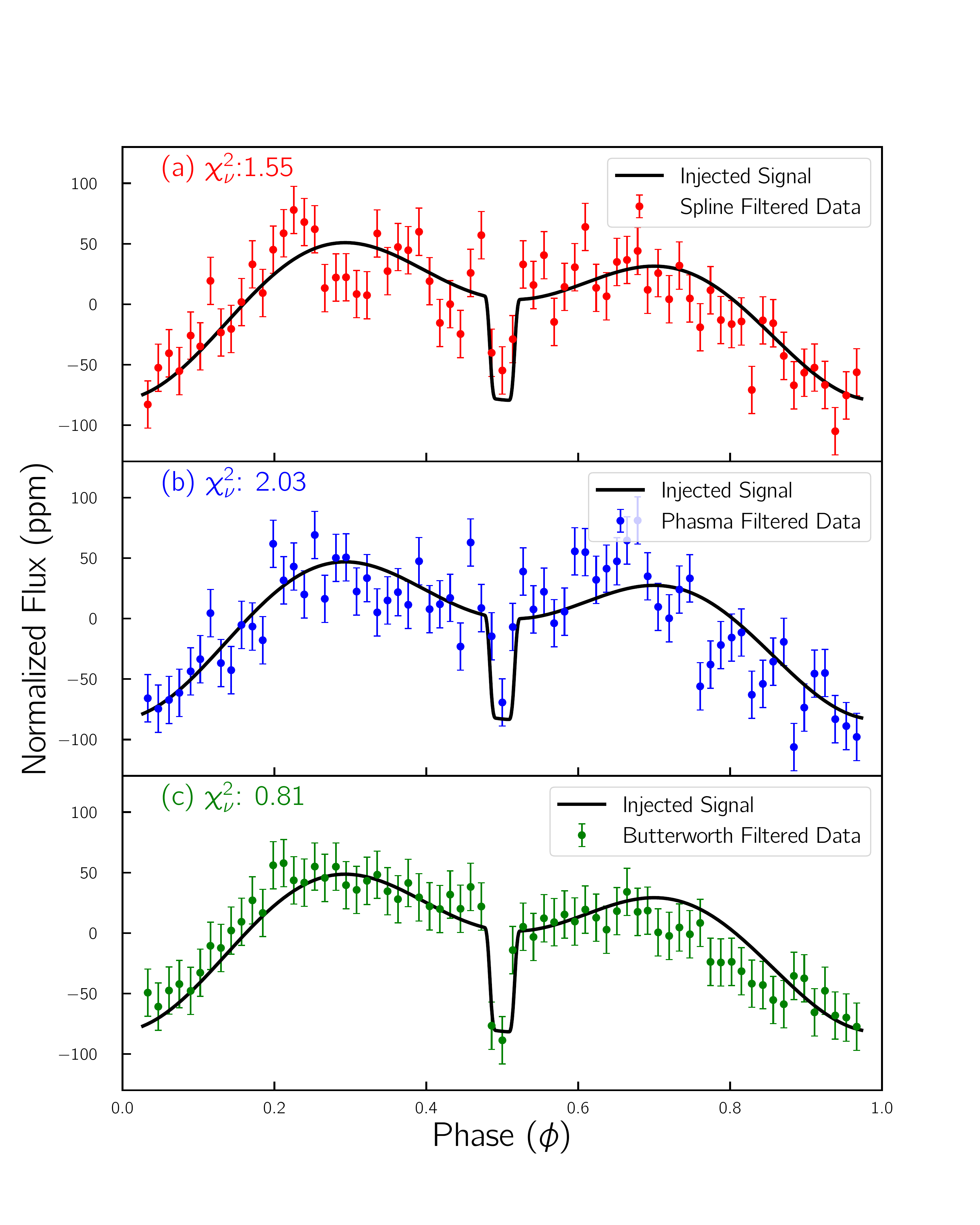}
\caption{\label{fig:threeflatteningmethod} A random case of injected signal retrieved through (a) Spline Filtering, (b) Phasma Filtering (c) Butterworth bandpass Filtering. The error bars  were scaled down accordingly to the bin-size from the calculated mean scatter.}
\end{figure}

\begin{figure}[ht]
\includegraphics[width=0.48\textwidth]{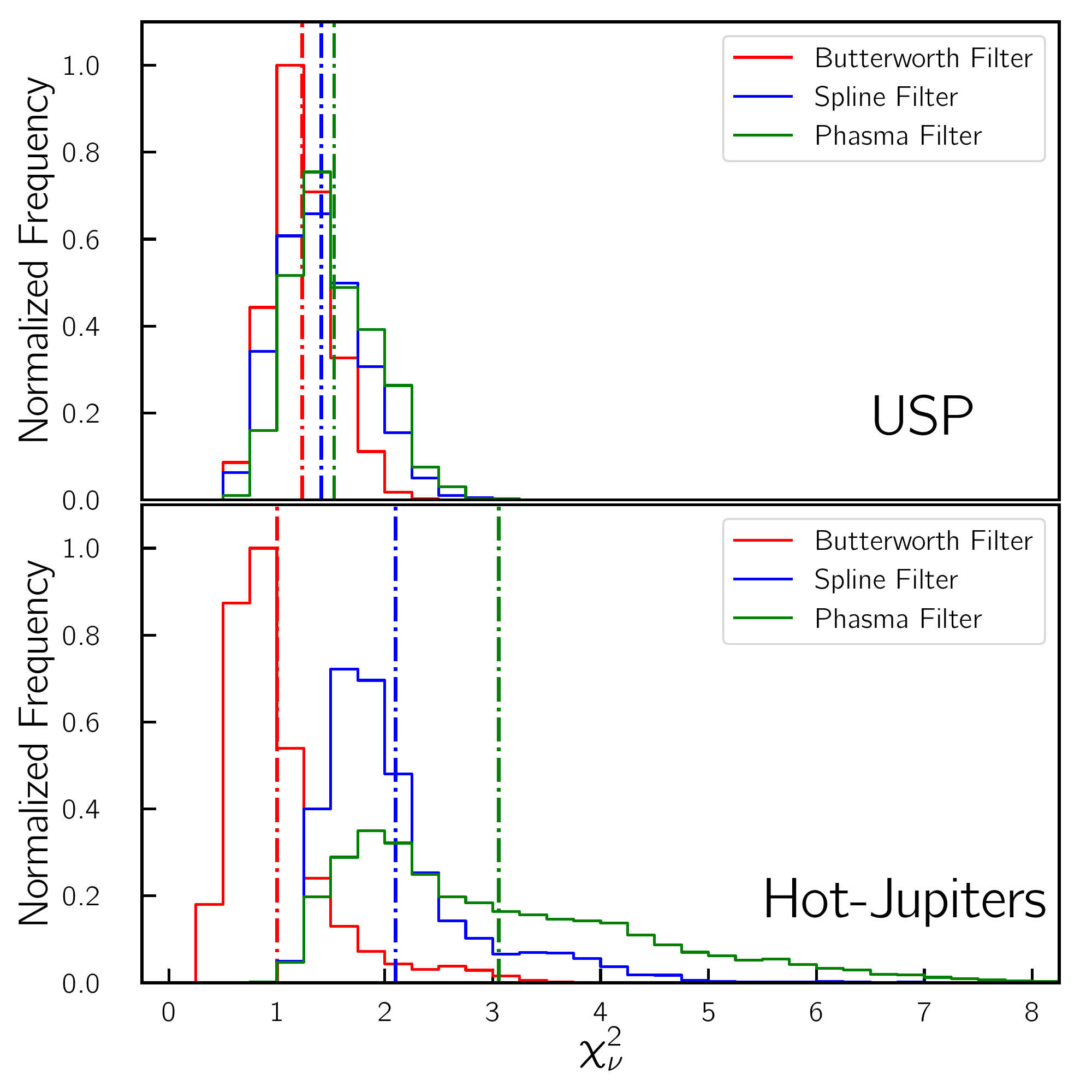}
\caption{\label{fig:retrievaltest} Top: Histogram of $\chi^2_\nu$ for the retrieved signal of the ultra-short period planets (USP). The performance of all three flattening methods are similar. Bottom: Histogram of $\chi^2_\nu$ for the retrieved signals observed for our samples of hot Jupiter targets. The mean value for shown with dotted lines  $\chi^2_\nu$ was 1.00 for Butterworth filters, 2.09 for Spline Filters, and 3.06 for Phasma Filters. Among the filtering techniques, we found the Butterworth filter statistically performed the best.}
\end{figure}

In order to extract the phase curve, the stellar continuum  has to be modeled out. A host of filtering techniques have been used in the past. Sometimes the stellar continuum exhibits little to no variation, therefore requiring  minimal pre-processing as in the case of TRES-2b \citep{barclay2012}. However, for most of \textit{K2} targets, filtering provides an opportunity not only to remove the stellar continuum, but also any uncorrected  systematics. Thus, we explored suites of filtering techniques available to us, among which we focused particularly on three: spline, phasma and Butterworth. Before performing all three filtering processes, we masked the region where the primary transit and secondary eclipse. This can lead to increased scatter around the region surrounding occultation. Below we discuss the three filtering techniques, whose performances were evaluated using $\chi_\nu^2$ as the primary metric:

\begin{enumerate}
\item \textbf{Spline Filtering}: Spline flattening is the most commonly used filtering technique for removing the stellar flux \citep{esteves2013, shporer2015,angerhausen2015, armstrong2016}. Use of different degrees of polynomial or knotting intervals are common depending on the planetary period as well as the ability of the spline to model the stellar continuum. Splines essentially act as a low pass filter, and have been successfully used in  a range of targets in the past. For this work, we adopt third degree polynomial knotted once every period of the planet.

\item \textbf{Phasma}: Phasma as a filtering method in the context of the phase curve was proposed in \citet{jansen2018}. Phasma in essence is a median filtering with window length set to the period of the planet. A similar implementation with a mean filter was also used. While easy to implement, phasma did not perform on par with other filtering techniques particularly among hot Jupiters (see \autoref{fig:threeflatteningmethod} and \autoref{fig:retrievaltest}).

\item \textbf{Frequency filtering}: Another method that has been used is harmonics based filtering \citep{quintana2013}. In this work, we use the sixth order Butterworth filter as implemented in \texttt{scipy}\footnote{\href{scipy.org}{scipy.org}} with a bandpass between half the planetary frequency and 3.5 times the planetary frequency. The limits of the bandpass were partially motivated to preserve the third harmonics \citep{esteves2015}, as well as decided through trial and error. Before filtering, we uniformly  and linearly interpolate the light curve after masking the transit and occultation points. Overall, the performance of Butterworth filter performance was superior among the three filtering techniques studied in detail, particularly among the hot Jupiters (see \autoref{fig:threeflatteningmethod}, and \autoref{fig:retrievaltest}). However, structural features such as ringing artifacts were more prominent in the residuals.  
\end{enumerate}

\begin{deluxetable}{lc}
\tablecaption{\label{table:InjectionTesttargets} Targets used for Signal Injection Test}
\tablehead{
\colhead{\textbf{Identifier}}&
\colhead{\textbf{EPIC IDs}}
}
\startdata
\textbf{K2-31b} & 203089855, 203526723, 203758400, \\
                &   204254456, 204529573\\
\textbf{HATS-9b} & 214576141, 214963629, 215327780\\
& 215496957, 216068131\\
\textbf{HATS-12b} & 215293111, 215310931 , 215517702\\
& 215594041, 215677034\\
\textbf{K2-107b} & 214402646, 215075353, 215542349\\
& 215771782, 215834357 \\
\textbf{K2-131b} & 201094825, 201094970, 201121210\\
& 201141186, 201164625\\
\textbf{K2-106b} & 220197918, 220205426, 220209263\\
& 220228282, 220249101
\enddata
\end{deluxetable}

For each system with detected phase curves, we download \texttt{K2SFF} and \texttt{EVEREST} detrended light curves of five target stars from the same campaign with similar magnitude and precision range after detrending. During the process of choosing the light curves for signal injection test, targets exhibiting strong short term modulation (i.e. less than 10 days), exhibiting intrinsic variability, or having obvious uncorrected systematics were deliberately avoided. The list of targets used for signal injection-retrieval test are reported in \autoref{table:InjectionTesttargets}. For the hot Jupiters among our targets, we injected the phase curve signals in the corresponding light curves considering geometric albedos between 0.01 to 0.66 at a step-size of 0.01 and masses between 0.25 and 7.0 $M_{\rm Jup}$ with a step-size of 0.25 $M_{\rm Jup}$ while using the reported transit parameters for the planets. We ran the subsequent light curves through our pipeline, and compared the final retrieved signal to the injected phase curve signal. The quick fits using Levenberg-Marquardt minimization showed that in most cases consistent parameters of albedo and mass to the injected signals can be retrieved. During these tests, we found the Butterworth filter outperforms both spline and phasma filtering among hot Jupiters (see \autoref{fig:threeflatteningmethod} and \autoref{fig:retrievaltest}). We also performed additional tests with phase offset signals, and Butterworth filter continued to outperform other two filters. For the ultra-short period super-Earths, we performed tests using parameters of K2-141, K2-131 and K2-106 for which we simulate phase curves with a reflective component with geometric albedo ranging between 0.01 to 0.8. For these latter set of planets, the performances of the different filtering techniques were comparable (see \autoref{fig:retrievaltest}).

Based on these tests, we have used Butterworth filtered light curves for all our hot Jupiter targets. The sixth order Butterworth filter allows us to isolate power from a  specific frequency band but it can sometimes lead to ringing effects in the residuals visible for some of our targets. Such artifacts were also observed in our injection retrieval test, however presence of these artifacts did not appear to affect the accuracy of the retrieved parameters. We explored the possibility of using narrower bandpass Butterworth filters, however, such implementations tended to decrease the accuracy of the retrieved parameters in our signal injection test. For the short period rocky planets (period less than a day) such as K2-141b, K2-131b or K2-106b, spline flattening light curves were used as its performance is comparable to Butterworth filter, and have already been shown to work robustly on a number of occasions.

Note that our injection-retrieval test has been heavily influenced by the  parameters of the systems where we discovered the phase curves. We performed our test using detrended lightcurves of those system that did not show strong stellar modulations or unusual artifacts. Thus, there are inherent limitations to the tests we performed, and the performance of the filtering techniques is likely to change in the presence of complex red noise. We have also limited ourselves to three filtering techniques in this paper, but there are a host of techniques available which we did not fully explore. For instance, some authors have pointed out the possibility of using Gaussian Processes \citep{serrano2018} for disentangling the planetary phase curves amid strong stellar modulations induced by stellar rotation. However, given the computational cost, the complexity of the model along with the possibility of overfitting the phase curves dissuaded us from diving too deep into this technique \citep{millholland2017}. In the future, we plan to explore a wider suite of filtering, and data analysis techniques which might allow us to improve on the process we introduce here.

\section{Models}
For all the targets where we detect phase curves, we simultaneously fit for the primary transit, secondary eclipse and phase curve. We set the period as noted in the NASA exoplanet archive to produce phase folded light curves. The simultaneous fit of the transit and the phase curves is primarily motivated to understand the cases with strong degeneracies among parameters as is the case of K2-31b. We initiate our MCMC model using the parameters reported in the discovery paper. For limb darkening, we use quadratic forms with uniform priors with range of 0.1 around the nearest values estimated by \citep{claret2011}. We similarly introduce priors in our MCMC for the scaled semi-major axis($a/R_*$) parameters to only accept values that yields the stellar density ($\rho_*$) within 5$\sigma$ of spectroscopically derived stellar density \citep{winn2010}:

\begin{equation}
    \rho_* + k^3 \rho_p = \frac{3 \pi  }{GP^2} \left( \frac{a}{R_*} \right)^3 ,
\label{eqn:stellardensity}    
\end{equation}
where $G$ is Gravitational constant, and $\rho_p$ is the planetary density. We ignore the contribution from  $k^3 \rho_p$ in our calculation. Additionally, we introduce a $T_0$ offset parameter, which allows for an offset in the time of the conjunction. Note that the final reported value of $T_0$ is calculated by appropriately combining this offset with the reported values in the original discovery paper.  

We use the \citet{mandel2002} formalism as implemented in \texttt{batman} \citep{kriedberg2015} in order to fit for the primary transit as well as the secondary eclipses. We supersample  our light curve by a factor of 15, and set the exposure time to 29.4 minutes. For the phase curves, we used Bayesian Information Criteria (BIC) to choose the best model  among three different models of phase curves: i) Thermal component with no significant nightside contribution (Model I) ii) Thermal component with  significant nightside contribution (Model II) iii) Thermal component with a phase offset (Model III). In Model I, the secondary eclipse depth is constrained as a function of the amplitude of the reflective and thermal component, which can lead  to compensated discovery of secondary eclipses. In Model II, the depth of the secondary eclipse is a free parameter without any priors. In Model III, we use the amplitude of the thermal component and phase offset as two free additional parameters compared to Model I. For all three models, we do not fit for phase variation during the transit, where we expect the transit to dominate the signal. We fit for a single mass parameter, which we refer to as the photometric mass ($M_{Phot}$), to model both ellipsoidal variation as well as Doppler effects. For all of our models, we have used non-eccentric models motivated by the original discovery papers. 

For exploring the parameter space, we used affine invariant MCMC implemented in \texttt{emcee} \citep{mackey2013} running for 25,000 steps with 50 walkers.  We use Gelman-Rubin statistics to ensure all of our MCMC converge. After the initial run, we compared the different models using Bayesian Information Criterion (BIC), and re-ran the best model for 50,000 step by initializing our parameters around the best obtained values in the previous runs. We built the posterior distribution to estimate the error after removing the first 25,000 step of the data, and use the rest to estimate the error of the fit parameters. From posterior distribution, we report the median and 1$\sigma$ confidence interval corresponding to 15.8$^{\rm th}$ percentile and 84.2$^{\rm th}$ percentile respectively. For some parameters such as planetary mass or equilibrium temperature, we propagate the error from the stellar parameters with the errors we estimate from the posterior distribution in our final reported parameters. 

\begin{figure*}[ht!]
\centering
\includegraphics[width=0.98\textwidth]{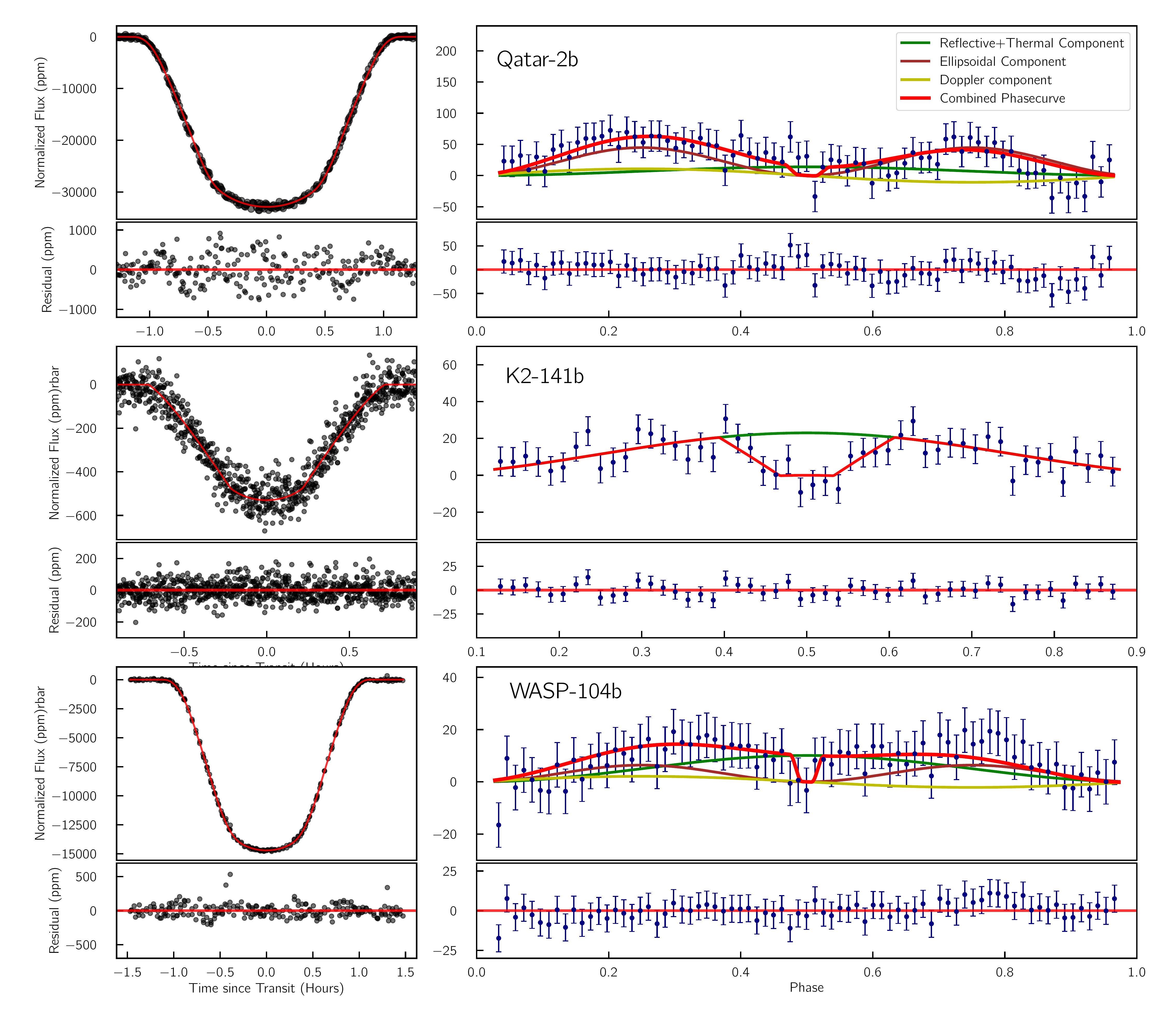}
\caption{\label{fig:ComboFigure0} \textbf{Top:} Best fit of phase curve model for Qatar-2b using \texttt{EVEREST} data exhibiting strong ellipsoidal and Doppler variation. \textbf{Middle:} Best fit f phase curve model of K2-141b showing strong thermal and reflective modulation. \textbf{Bottom} Phase curve of WASP-104b observed in \texttt{K2SFF} detrended light curve exhibiting prominent ellipsoidal and Doppler effects. The fit parameters for all three targets are presented in \autoref{table:phasefit_param0}.} 
\end{figure*}

For all our models we consider circular orbits and expect the secondary eclipse to occur exactly at the half phase.  We ignore the R\o mer delay, and  the effects of eccentricity. These choices were motivated by the precision of the data as well as the reporting of no significant eccentricities in the  discovery papers. As for the temperature of the planet, we report the dayside temperature by numerically solving the following equation:
\begin{equation}
    \Delta = A_g \left(\frac{R_p/R_*}{a/R_*}\right)^2 + \left(\frac{R_p}{R_*}\right)^2 \frac{\int B(T_{Day}) R(\lambda) d \lambda}{\int B(T_{e\!f\!f}) R(\lambda)d \lambda},
\label{eqn:delta_ag}    
\end{equation}
where $\Delta$ is the secondary eclipse depth, $R_p/R_*$ is the scaled radius, $a/R_*$ is the scaled semi-major axis, and $B$ is Planck function which is convolved with the \textit{Kepler} response function $R(\lambda)$. We solve \autoref{eqn:delta_ag} for the geometric albedo ($A_g$), and assume the day-side contribution is a function of the geometric albedo using aforementioned \citet{lopez2007} formalism introduced in  \autoref{eqn:lopez2007}. In order to report the dayside temperature, we set the re-radiation factor to 1/2  and assume A$_B$ = 3/2A$_g$ unless otherwise stated. Note this simple relation is not valid for planets in our solar system, and usually is likely to overestimate the value of the Bond Albedo \citep{dyudina2016}. We also require the secondary eclipse depth to be strictly greater than or equal to the amplitude of the reflective and thermal component at phase 0.5. Similarly, the equilibrium temperature is calculated by setting the re-radiation factor to be 1/4.

\section{Results}

\subsection{Pre-Selected Targets}

Prior to our study, three of K2 targets had reported phase curves: Qatar-2b \citep{dai2017b}, K2-141b \citep{malavolta2018}, and WASP-104b \citep{mocnik2018}. We re-run these objects through our pipeline, and fit the phase folded data and fit the phase curve models. For QATAR-2b, we obtained ellipsoidal variation amplitude of 22.6$^{+2.6}_{-2.5}$ ppm against 15.4$\pm$4.8 ppm, and Doppler modulation amplitude of 10.8$^{+1.3}_{-1.2}$ ppm against 14.6 $\pm$ 5.1 ppm reported in \citet{dai2017b}. This led to estimation of the photometric mass of 3.27$^{+0.40}_{-0.41}$ \mj consistent within 2$\sigma$  with the radial velocity mass of 2.487$\pm$0.089 \mj \citep{bryan2012}. The reflective and the thermal component in the phase curve constrain the geometric albedo with 3$\sigma$ confidence at 0.03 under the assumption $A_g = 3/2A_B$.  For K2-141b, we use the use the same detrended lightcurve used in \citet{malavolta2018}, and the amplitude of the phase curve we find 11.8$\pm$1.5 ppm is consistent to their finding of occultation depth of 23 $\pm$ 4 ppm. Under the assumption $A_g = A_B$, and numerically solving \autoref{eqn:delta_ag}, we estimate the geometric albedo of 0.205$^{+0.059}_{- 0.077}$. Similarly, we detect phase curve signal in WASP-104b using \texttt{K2SFF} detrended light curve as had been reported in \citet{mocnik2018}. We find the reflective and thermal component  at 5.1$\pm$1.0 against theirs reporting at 4.8$\pm$2.1 ppm, ellipsoidal variation amplitude at 3.21$\pm$0.83 against theirs reporting at 6.9$\pm$2.2 ppm, and Doppler effect amplitude at  2.13$\pm$0.55 against theirs reporting at 4.2$\pm$1.9 ppm. Also the photometric mass we obtain 0.99$\pm$0.26 \mj is consistent within 1.5$\sigma$  with the radial velocity mass 1.272$\pm$0.047 \citep{smith2014}.

Thus, our results for all three planets are largely consistent to the previous phase curves analyses. The fits using the best parameters for our targets are presented in \autoref{fig:ComboFigure0}, and the corresponding fit parameters are presented in \autoref{table:phasefit_param0}. The light curves used in the analysis, as well as filtered light curves has been presented in the Appendix.

\begin{deluxetable*}{lcccc}
\tablecaption{\label{table:phasefit_param0} Stellar and Planetary Parameters for Qatar-2b, K2-141b, and WASP-104b}
\tablehead{
\colhead{Parameter} &
\colhead{Unit}&
\colhead{QATAR-2b}&
\colhead{K2-141b}&
\colhead{WASP-104b}
}
\startdata
\multicolumn{2}{l}{\textbf{Stellar Parameters}}\\
$M_*$ & $M_\odot$ & 0.74$\pm$0.04$^{a}$ & 0.708$\pm$0.028$^{b}$ &  1.02$\pm$0.09$^{c}$\\
$R_*$  & $R_\odot$ & 0.713$\pm$0.018$^{a}$& 0.681$\pm$0.018$^{b}$& 0.93$\pm$0.23$^{c}$\\
$T_{e\!f\!f}$ & K & 4645$\pm$50 $^{a}$ & 4599$\pm$79$^{b}$\ & 5450$\pm$130$^{c}$\\
$[$Fe/H$]$  & dex & -0.02$\pm$0.08$^{d}$& -0.06$^{+0.08}_{-0.10}$$^{b}$ & -0.09$\pm$0.09$^{c}$ \\
log $g_*$  & cgs & 4.601$\pm$0.018$^{a}$& 4.62$^{+0.02}_{-0.03}$$^{b}$ & 4.5$\pm$0.2$^{c}$ \\
$u$ & -& 0.7165$^{e}$& 0.7194$^{e}$ & 0.6459$^{e}$\\
$g$ & -& 0.5434$^{e}$& 0.5452$^{e}$ & 0.4292$^{e}$\\
\hline
\textbf{Pipeline}& - & \texttt{EVEREST} & \texttt{K2SFF}$^{b}$ & \texttt{K2SFF}\\
\hline
\multicolumn{2}{l}{\textbf{Orbital Parameters}}\\
Period  & Days & 1.337116553$\pm$0.000000044$^{f}$& 0.2803244$\pm$0.0000015$^{b}$ & 1.75540636$\pm$0.00000014$^{g}$\\
$T_0$ - 2450000 & BJD & 5617.5816109$\pm$0.0000087 & 7744.071542$^{+0.000071}_{-0.000068}$ &  7935.0702204$\pm$0.0000078\\
$R_p/R_*$ & - &  0.16526$^{+0.00010}_{-0.00009}$ & 0.02084$^{+0.0005}_{-0.0002}$& 0.12041$\pm$0.00026 \\
$a/R_*$ & - & 6.6769$^{+0.006}_{-0.013}$& 2.30$^{+0.05}_{-0.15}$ & 6.732$^{+0.041}_{-0.044}$ \\
$b$  & - &  0.036$^{+0.038}_{-0.026}$ & 0.23$^{+(0.22}_{-0.16}$ & 0.7115$^{+0.0063}_{-0.0059}$\\
Inclination& Deg &  89.69$^{+ 0.22}_{- 0.33}$ &  84.2 $^{+ 4.1}_{-6.3}$ & 83.933$^{+0.087}_{-0.093}$\\
$e$ &-& 0 (assumed)& 0 (assumed) & 0 (assumed) \\
$\omega$ & Deg & 90 (assumed)& 90 (assumed) & 90 (assumed)\\ 
$u_1$  & - & 0.5488$^{+0.0016}_{-0.0008}$ & 0.639$^{+0.057}_{-0.051}$ & 0.422$^{+0.017}_{-0.015}$\\
$u_2$  & - & -0.0051$^{+0.0043}_{-0.0020}$ & 0.074$^{+(0.073}_{-0.061}$ & 0.135$^{+0.023}_{-0.015}$\\
\hline
\multicolumn{2}{l}{\textbf{Phase Curve Parameters}}\\
$A_{Re\!f+T\!h}$ & ppm & 6.9$^{+3.1}_{-3.0}$ & 11.8$\pm$1.5 & 5.1$\pm$1.0\\
$A_{Ell}$ & ppm &  22.6 $^{+2.6}_{- 2.5}$ & -- & 3.21$\pm$0.83\\
$A_{Dop}$ & ppm & 10.8$^{+1.3}_{-1.2}$	 & -- & 2.13$\pm$0.55\\
$T_{Day}$ & K & 1711 $\pm$ 24 & 2406$^{+ 144}_{- 76}$
 & 1698 $\pm$24\\
$T_{eq}$ & K& 1434 $\pm$ 20 & 1984$^{+108}_{-59}$ & 1422 $\pm$ 20\\
$A_g$ & - & $<$0.03 (3$\sigma$) & 0.205$^{+0.059}_{- 0.077}$
 & 0.0211 $\pm$ 0.0068\\
$M_{RV}$ & $M_{\mathrm{Jup}}$ & 2.487$\pm$0.086$^{a}$  & 0.016$\pm$0.0013$^{b}$ & 1.272$\pm$0.047$^{c}$\\
$M_{Phot}$ & $M_{\mathrm{Jup}}$ & 3.27$^{+0.40}_{-0.41}$ & -- & 0.99$\pm$0.26
\enddata
\tablenotetext{a}{Adopted from \citet{bryan2012}}
\tablenotetext{b}{Adopted from \citet{malavolta2018}}
\tablenotetext{c}{Adopted from \citet{smith2014}}
\tablenotetext{d}{Adopted from \citet{maxted2015}}
\tablenotetext{e}{Linear limb-darkening ($u$) and gravity-darkening ($g$) coefficients interpolated from \citet{claret2011}}
\tablenotetext{f}{Adopted from \citet{dai2017b}}
\tablenotetext{g}{Adopted from \citet{mocnik2018}}
\end{deluxetable*}

\subsection{New Discoveries}
From the remaining 49 targets, we discovered phase curves among 6 of planets. Once discovered, we fitted three standard models as discussed under $\S$5. The unfiltered light curve, the outliers, as well as the filtered light curves from the process are added presented under the Appendix. Below we present and discuss each of our targets in detail:

\begin{deluxetable*}{lccccc}
\tablecaption{\label{table:phasefit_param} Stellar and Planetary Parameters for K2-31b, HATS-9b, HATS-12b, and K2-107b}
\tablehead{
\colhead{Parameter} &
\colhead{Unit}&
\colhead{K2-31b}&
\colhead{HATS-9b}&
\colhead{HATS-12b}&
\colhead{K2-107b}
}
\startdata
\multicolumn{2}{l}{\textbf{Stellar Parameters}}\\
$M_*$ & $M_\odot$ & 0.91$\pm$0.06$^{a}$ & 1.030$\pm$0.039$^{b}$ & 1.489$\pm$0.071$^{c}$ & 1.30$\pm$0.14$^{d}$\\
$R_*$  & $R_\odot$ & 0.78$\pm$0.07$^{a}$& 1.503$^{+0.101}_{-0.043}$$^{b}$ & 2.21$\pm$0.21$^{c}$ & 1.78$\pm$0.16$^{d}$\\
$T_{e\!f\!f}$ & K & 5280$\pm$70$^{a}$& 5366 $\pm$70$^{b}$\ & 6408$\pm$75$^{c}$ & 6030$\pm$120$^{d}$\\
$[$Fe/H$]$  & dex & 0.08$\pm$0.05$^{a}$& 0.340$\pm$0.050$^{b}$ & -0.100$\pm$0.040$^{c}$ & 0.10$\pm$0.10$^{d}$\\
log $g_*$  & cgs &4.60$\pm$0.07$^{a}$& 4.095$\pm$0.038$^{b}$ & 3.923$\pm$0.065$^{c}$ & 4.07$\pm$0.10$^{d}$\\
$u$ & -& 0.6554$^{e}$& 0.5467$^{e}$ & 0.5317$^{e}$ & 0.5723$^{e}$\\
$g$ & -& 0.4611$^{e}$& 0.3200$^{e}$ & 0.2751$^{e}$ & 0.3280$^{e}$\\
\hline
\textbf{Pipeline}&-& \texttt{EVEREST} & \texttt{K2SFF} 
& \texttt{K2SFF} & \texttt{EVEREST}\\
\hline
\multicolumn{2}{l}{\textbf{Orbital Parameters}}\\
Period  & Days & 1.257850$\pm$0.000002$^{a}$& 1.9153073$\pm$0.000005$^{b}$ & 3.142833 $\pm$0.000011$^{c}$ & 3.31392$\pm$0.00002$^{d}$\\
$T_0$ - 2450000 & BJD & 2358.709367$\pm$0.000010& 6124.258934$^{+0.000032}_{-0.000033}$ &  6798.955644$\pm$0.000075 & 6928.059202$\pm$0.000055\\
$R_p/R_*$ & - &  0.168$^{+0.042}_{-0.023}$& 0.08414$^{+0.00014}_{-0.00012}$& 0.06048$^{+0.00056}_{-0.00042}$ & 0.08335$^{+0.00023}_{-0.00026}$\\
$a/R_*$ & - & 5.66$^{+0.10}_{-0.09}$&  4.556$^{+0.011}_{-0.026}$ & 5.47$^{+0.19}_{-0.24}$ & 5.890$^{+0.082}_{-0.075}$\\
$b$  & - &  1.022$^{+0.053}_{-0.031}$ &  0.065$^{+0.065}_{-0.045}$ & 0.29$^{+0.12}_{-0.16}$ & 0.7925$^{+0.0074}_{-0.0082}$\\
Inclination& Deg & 79.61$^{+0.51}_{-0.74}$ &  89.29 $^{+0.55}_{-0.82}$ & 87.0$^{+1.8 }_{-1.5}$  & 82.27$^{+ 0.18}_{- 0.17}$\\
$e$ &-& 0 (assumed)& 0 (assumed) & 0 (assumed) & 0 (assumed)\\
$\omega$ & Deg & 90 (assumed)& 90 (assumed) & 90 (assumed) & 90 (assumed)\\
$u_1$  & - & 0.560$^{+0.029}_{-0.051}$ &  0.5300$^{+0.0072}_{-0.0049}$ & 0.3227$^{+0.0091}_{-0.0053}$ & 0.437$^{+0.027}_{-0.019}$\\
$u_2$  & - & 0.261$^{+0.039}_{-0.062}$ & -0.012$^{+0.017}_{-0.016}$  & 0.236$^{+0.013}_{-0.006}$ & 0.100$^{+0.030}_{-0.021}$\\
\hline
\multicolumn{2}{l}{\textbf{Phase Curve Parameters}}\\
$A_{Re\!f+T\!h}$ & ppm & 12.27$^{+0.85}_{-0.83}$ &11.6$^{+2.3}_{-2.4}$ & 7.5$\pm1.9$ & 12.8$^{+1.9}_{-2.0}$ \\
$A_{Ell}$ & ppm & 17.07$^{+ 0.77}_{-0.75}$ &  14.7$\pm$ 2.3& 10.2$\pm$1.8  & 11.6 $^{+ 1.8}_{-1.9}$\\
$A_{Dop}$ & ppm & 5.55$^{+ 0.33}_{- 0.30}$ & 2.08 $\pm$ 0.33 &  2.53$^{+ 0.54 }_{-0.50 }$ &  3.03$^{+ 0.51}_{-0.50}$\\
$T_{Day}$ & K& 1860$\pm$35 &  2100$\pm$29& 2240$^{+82}_{-68}$ & 2005$^{+ 26}_{-27}$\\
$T_{eq}$ & K& 1554$\pm$29 & 1751$\pm$24  & 1849.$^{+61}_{-52}$ & 1669$^{+20}_{-21}$\\
$A_g$ & - & $<$0.047 (3$\sigma$) & 0.027$^{+ 0.015}_{-0.017}$ & 0.07$\pm 0.04$ &  0.102$\pm$0.023\\
$M_{RV}$ & $M_{\mathrm{Jup}}$ & 1.774$\pm$0.079$^{a}$& 0.837$\pm$0.029$^{b}$ &  2.38$\pm$ 0.11$^{c}$ &  0.84$\pm$0.08$^{d}$\\
$M_{Phot}$ & $M_{\mathrm{Jup}}$ & 2.09$\pm0.18$  & 0.98$\pm0.16$ & 2.13$^{+0.46}_{-0.43}$& 1.57$^{+0.26}_{-0.27}$\\
\enddata
\tablenotetext{a}{Adopted from \citet{grziwa2016}}
\tablenotetext{b}{Adopted from \citet{brahm2015}}
\tablenotetext{c}{Adopted from \citet{rabus2016}}
\tablenotetext{d}{Adopted from \citet{eigmuller2017}}
\tablenotetext{e}{Linear limb-darkening ($u$) and gravity-darkening ($g$) coefficients interpolated from \citet{claret2011}}
\end{deluxetable*}

\subsubsection{K2-31b} 
Given the short period, high mass and large scaled radius, K2-31b was the best candidate among \textit{K2} discovered planets to have a detectable phase curve. Our analysis of the data indeed shows the presence of a robust phase curve with all of the components present. K2-31b is a grazing hot Jupiter with mass of $\sim$1.8 M$_{\mathrm{Jup}}$ discovered by \citet{grziwa2016}. Due to the grazing nature of the transit, which introduces a degeneracy between scaled radius and inclination, the radius of the planet has not very precisely determined. Our MCMC yielded the scaled radius of the planet of 0.168$^{+0.042}_{-0.023}$, which assuming a host star of radius 0.78 $R_\odot$ translates into physical radius of 1.28$^{+0.32}_{-0.17}~R_\mathrm{Jup}$.

\begin{figure*}[ht!]
\centering
\includegraphics[width=0.98\textwidth]{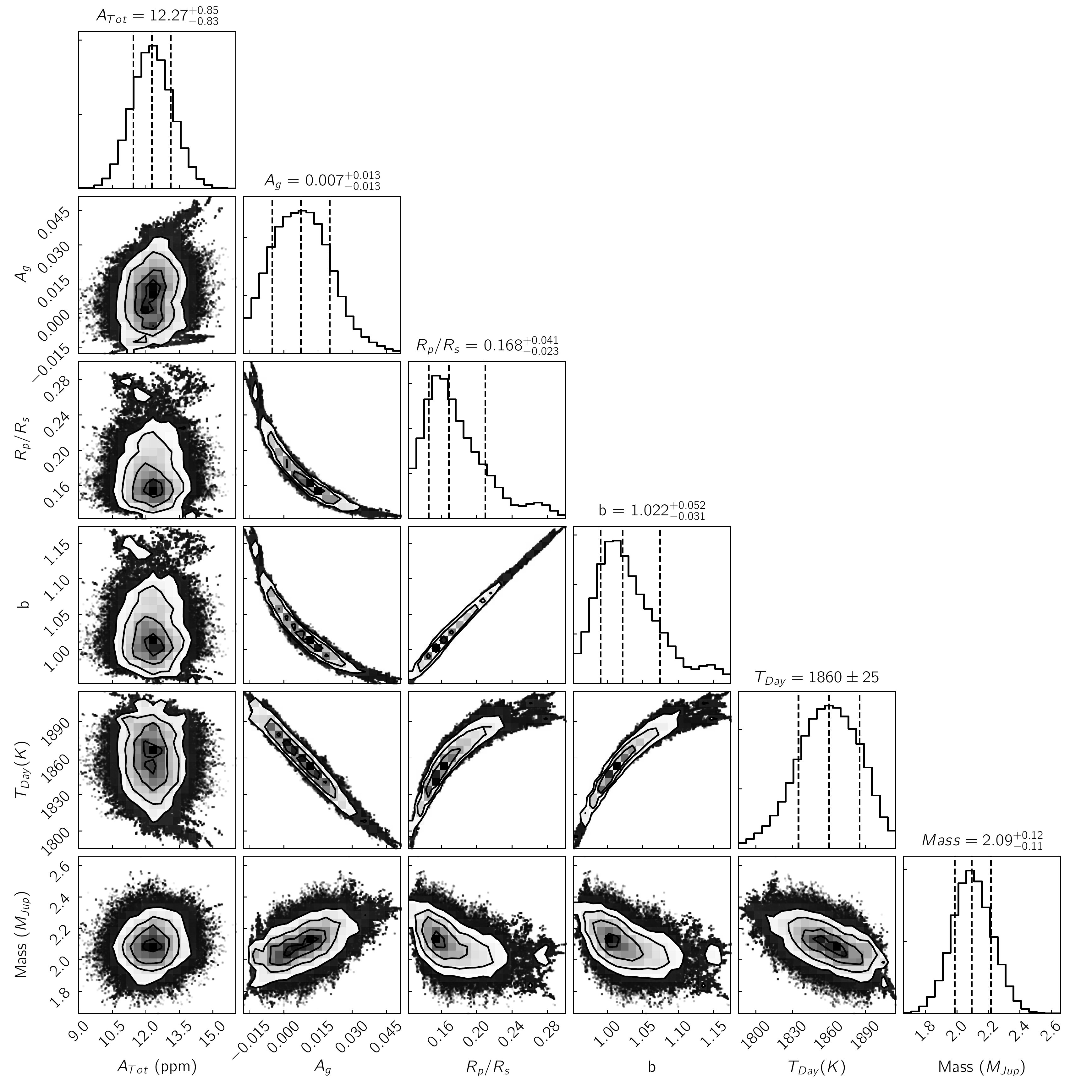}
\caption{\label{fig:k2_31_cornerplot} Corner plot showing the posterior distribution and co-variance among different fit parameters of K2-31b. A strong degeneracy is present among the scaled radius ($Rp/R_*$), geometric albedo ($A_g$), impact parameter ($b$), and day-side temperature ($T_{Day}$), while the mass of the planet exhibits minimal correlation with most of the parameters. The errors from the stellar parameters have not  been propagated.} 
\end{figure*}
For our analysis, we use the \texttt{EVEREST} detrended light curve and obtain a photometric mass ($M_{phot}$) of  2.09$\pm0.18$ $M_{\mathrm{Jup}}$ consistent within 2$\sigma$ to the RV-based mass ($M_{RV}$) of 1.774$\pm$0.079 $M_{\mathrm{Jup}}$ reported in the discovery paper. \citet{grziwa2016} points out that due to its grazing nature which is often interpreted as eclipsing binaries, planets like K2-31b are usually neglected for RV follow-up. We show here that a target's planetary nature can be revealed by estimating mass through the optical phase curve. In the discovery paper, \citet{grziwa2016} constrains the upper limit on the geometric albedo of K2-31b at 0.40, due to the absence of any visible secondary eclipse. By fitting for the whole phase curve, we are able to constrain the geometric albedo to 0.047 at 3$\sigma$ confidence, a  dark planet even by the standard of hot Jupiters \citep{esteves2015, angerhausen2015}. Yet, the grazing nature of the transit leads to strong degeneracy among parameters (see \autoref{fig:k2_31_cornerplot}), as well as the underlying assumptions for the thermal emission and the Bond albedo leaves room for unaccounted systematic errors for such estimation. The degeneracy among different parameters, however is not as strong for the observed photometric mass. A ringing feature appears to present in the residual, although it is not a very prominent one.

\begin{figure*}[ht!]
\centering
\includegraphics[width=1.02\textwidth]{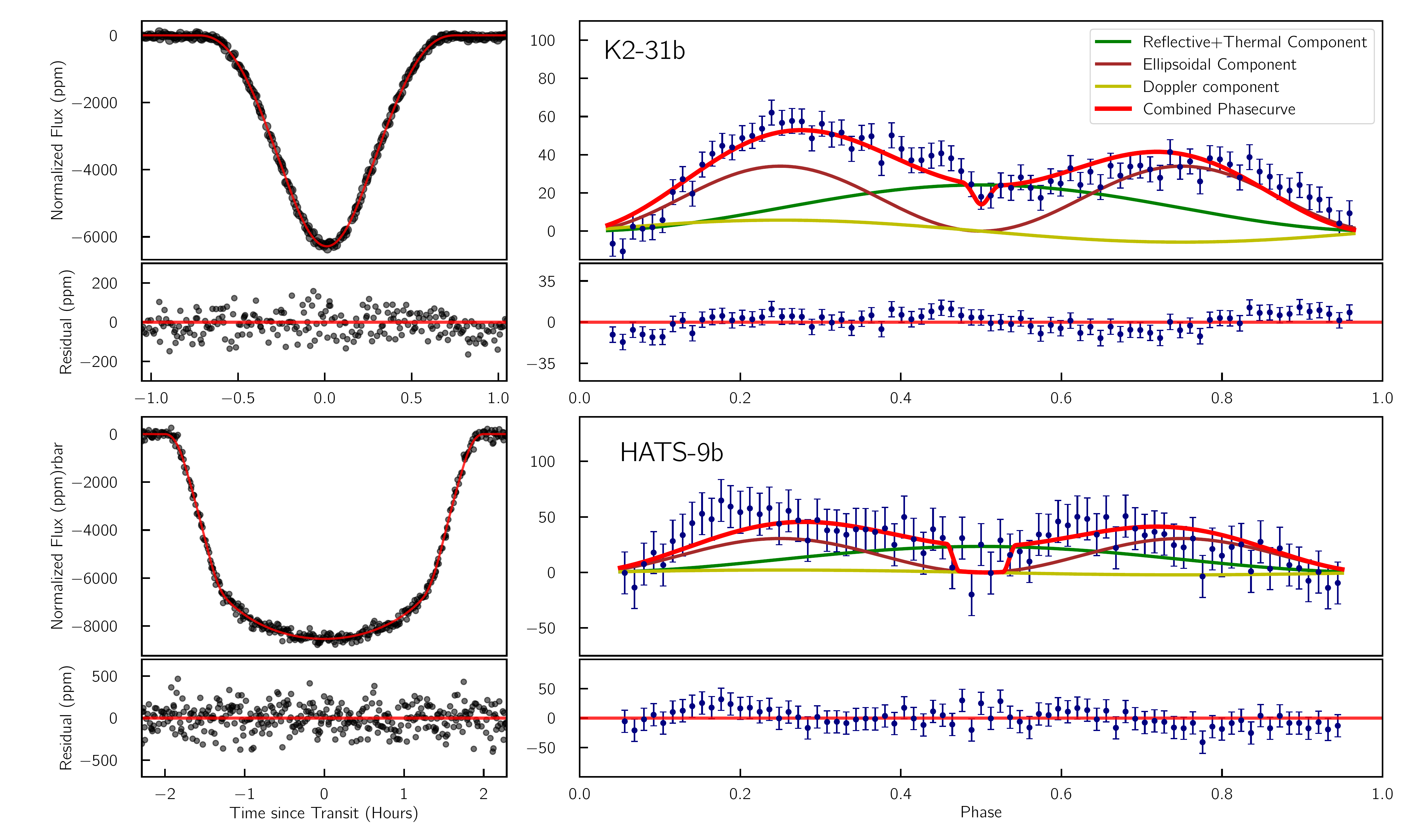}
\caption{\label{fig:ComboFigure1} Upper-left: Transit fit using the best fit parameter for K2-31b and the residual. Upper-right: Phase curve signal using the best obtained parameter for K2-31b with different components. The data was binned to have a total number of bins of 75, thus each bin size corresponds to 0.37 hours. Lower-left: Transit fit using the best fit parameter for HATS-12b and the residual. Lower-right: Phase curve signal using the best set of visualization in HATS-9b with different components and its residual. The data was binned to have a total number of bins of 75, thus each bin size corresponds to 0.55 hours.} 
\end{figure*}

\subsubsection{HATS-9}
HATS-9b is a hot Jupiter that was discovered in the Campaign 7 \textit{K2} field  \citep{brahm2015}. \citet{bayliss2018} updated the system parameters using the \textit{K2} lightcurve, however did not report the phase curve. We use \texttt{K2SFF} detrended light curve for this particular analysis given its precision, and in examining the light curve for the potential phase curve, we detect the ellipsoidal variation for HATS-9b which yielded a photometric mass of 0.98$\pm0.16$ M$_{\mathrm{Jup}}$ consistent within 1$\sigma$ to the reported RV mass of 0.837 $\pm0.029$ M$_{\mathrm{Jup}}$. Fitting for the reflective and the thermal component of the phase curve yielded the geometric albedo of 0.027$^{+ 0.015}_{-0.017}$. The fit using the best parameters is shown in \autoref{fig:ComboFigure1}, and the corresponding parameters are reported in  \autoref{table:phasefit_param}. Ringing effects appears to be more prominent among the residual of HATS-9b.

\begin{figure*}[ht!]
\centering
\includegraphics[width=1.02\textwidth]{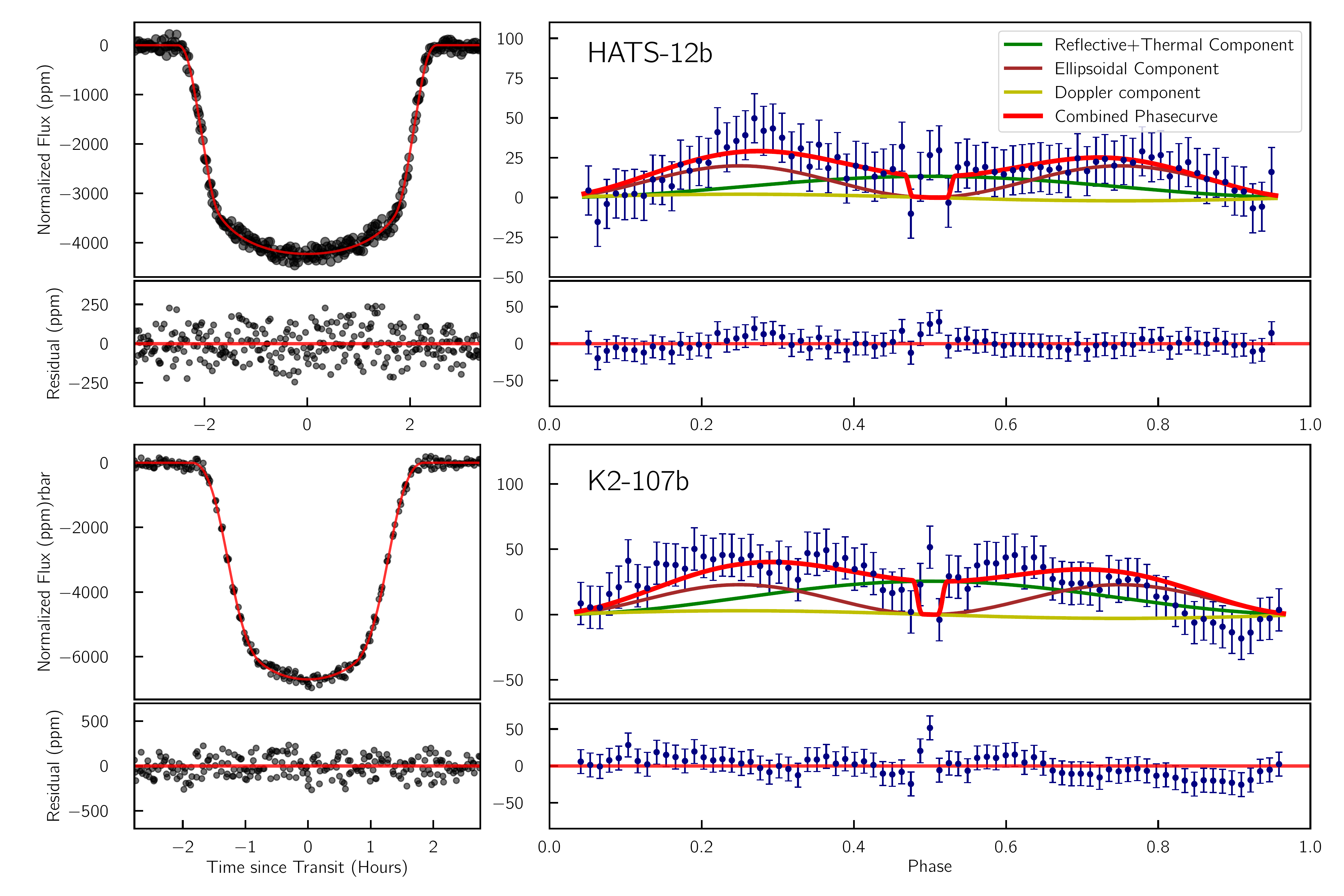}
\caption{\label{fig:ComboFigure2} Upper-left: Transit fit using the best fit parameter for HATS-12b and its residual. Upper-right: Phase curve signal using the best obtained parameter for HATS-12b with different components. The data was binned to have a total number of bins of 75, thus each bin size corresponded to 0.91 hours. Lower-left: Transit fit using the best fit parameter obtained for K2-107b and its residual. Lower-right: Phase curve signal using the best set of visualization in K2-107b with different components and its residual. The data was binned to to have a total number of bins of 75, thus each bin size corresponds to 0.99 hours.} 
\end{figure*}

\subsubsection{HATS-12b}
HATS-12b  \citep{rabus2016} was another hot Jupiter discovered in the
\textit{K2} Campaign 7 field.  An abrupt jump in the data was observed in both detrended light curves around BJD - 2457333.1 (see \autoref{fig:anomaly}), possibly due to a change in the pixel responsivity \citep{jenkins2010}. We use \texttt{K2SFF} detrended light curves for the analysis, and corrected for the jump by using a linear regression at the break point.  Marked as one of the promising targets using our SNR metric, we see distinct phase curve emerge in the phase folded light curve.  The phase curves exhibits particularly prominent ellipsoidal variation, fitting for which leads to a mass of 2.13$^{+0.46}_{-0.43}$ $M_{\mathrm{Jup}}$ consistent within 1$\sigma$ to the reported RV mass of 2.38$\pm 0.11$ $M_{\mathrm{Jup}}$ \citep{rabus2016}. The good agreement comes with a little surprise given HATS-12 has a stellar mass of 1.489$\pm$ 0.071 $M_\odot$, which lies above 1.4 $M_\odot$, a threshold beyond which the tidal-equilibrium approximation assumed for calculating the ellipsoidal variation amplitude may not strictly hold \citep{pfahl2008}.   We find the geometric albedo is 0.07$\pm 0.04$, typical for hot Jupiters. The fit using the best set of parameters is shown in \autoref{fig:ComboFigure2}, and the corresponding parameters are reported in  \autoref{table:phasefit_param}.  Note the secondary eclipse primarily comes as a constraint from the use of Model I, which is favored among three models using BIC.

\subsubsection{K2-107b}
K2-107b was reported in \citet{eigmuller2017}  where using high-resolution imaging, a few nearby companions were detected. However, the dilution due to these companions is only 0.5$\pm$0.1\% of the \textit{K2} aperture flux and therefore was ignored in this analysis. A few nearby companions were detected with high resolution imaging, which are positioned in the \textit{K2} aperture, however the combined dilution factor correction due to these companion is 0.005$\pm$0.001, therefore negligible for the calculation we are considering. We use the \texttt{EVEREST} detrended light curve for the fitting purposes, which yields a photometric mass of   1.57$^{+0.26}_{-0.27}$ M$_{\mathrm{Jup}}$, which is within 3$\sigma$ of  RV mass reported at  0.84$\pm$0.08 M$_{\mathrm{Jup}}$. The estimated geometric albedo is 0.102$\pm0.023$.  The fit of K2-107b is shown in \autoref{fig:ComboFigure2}, and the parameters are reported in  \autoref{table:phasefit_param}. Some structural features are present in the residuals potentially because of the ringing effects from filtering process. Like in HATS-12b, the secondary eclipse is a constraint from the use of Model I. The larger scatter of the data points during the secondary eclipse potentially results from the masking of these points during the filtering process.

\subsubsection{K2-131b}
\begin{figure*}[ht!]
\centering
\includegraphics[width=1.02\textwidth]{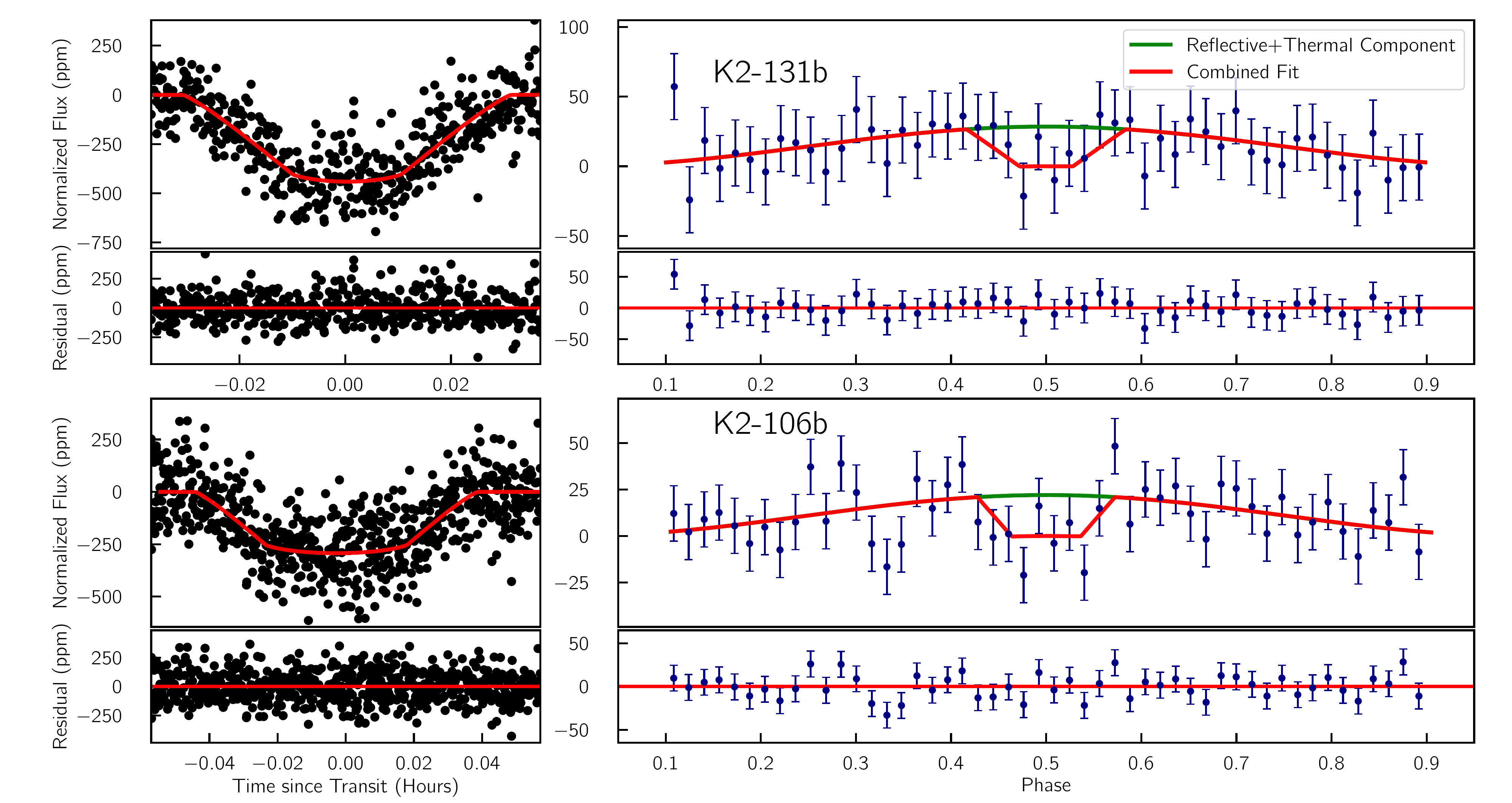}
\caption{\label{fig:ComboFigure3} Upper-right: Transit fit using the best parameter for K2-131b. Upper-left: Phase curve modulation showing reflective and thermal component in K2-131b with a secondary eclipse depth at 25.4$\pm$8.2 ppm. A total of 50 bins are used which corresponds to bin size of 0.14 hours. Lower-right: Transit fit using the best parameter for K2-106b. Lower-left: Phase curve modulation showing reflection and thermal component modulation in the folded light curve of K2-106b with a secondary eclipse depth at  23.5$^{+4.9}_{-5.1}$ ppm. A total of 50 bins are used which corresponds to bin size of 0.22 hours.}
\end{figure*}

K2-131b is an ultra-short period planet with a period of 0.3693 days reported in \citet{dai2017a}. For this work, we use the light curve from \texttt{EVEREST} pipeline, and only from the second half of Campaign 10  as the first half of the data shows strong systematic effects. In the phase folded data, we detect the secondary eclipse at 25.4$\pm$8.2 ppm. We only fit for reflective and thermal component, as the ellipsoidal and Doppler beaming signal from the small planets is not detectable. Unlike for the hot Jupiters in our lists, we use an altered relation between the geometric albedo and Bond Albedo as $A_B = A_g$. The modified relation allows for values greater than 0.66 for geometrical albedo, and such deviation from the traditional Lambertian relation ($A_B$ = $\frac{3}{2} A_g$) is expected as it tends to overestimate the Bond Albedo \citep{dyudina2016}. The fits showing the best obtained parameters are shown in \autoref{fig:ComboFigure3}, and the parameters are reported in \autoref{table:secondaryeclp_param}. Note the geometric albedo of 0.27$^{+0.31}_{-0.27}$ was estimated for K2-131b. 

\subsubsection{K2-106b}
K2-106 is a multi-planetary system with an ultra short period planet with period of 0.57133 days. It was first identified as a candidate in \citet{adams2017}, and  subsequent RV campaigns verified the planetary nature of the signal with mass reported in \citet{guenther2017} and \citet{sinukoff2017}. For our analysis, we adopted the values from the latter. We detect the secondary eclipse  at 23.5$^{+4.9}_{-5.1}$ ppm. Reflective and thermal component constitute the dominant part of the phase curve (see \autoref{fig:ComboFigure3}), and like K2-131b, we do not fit for either ellipsoidal variation or Doppler beaming due to negligible expected contributions. In order to estimate the temperature, we similarly use the modified relation i.e. $A_B = A_g$, which yielded a geometric albedo of 0.62$^{+0.22}_{-0.34}$. The fit itself is shown in \autoref{fig:ComboFigure3}, and the corresponding fit parameters are reported in \autoref{table:secondaryeclp_param}.

\section{Secondary Eclipse}
\label{sec:SecEclip}

\begin{figure}[ht!]
\centering
\includegraphics[width=0.49\textwidth]{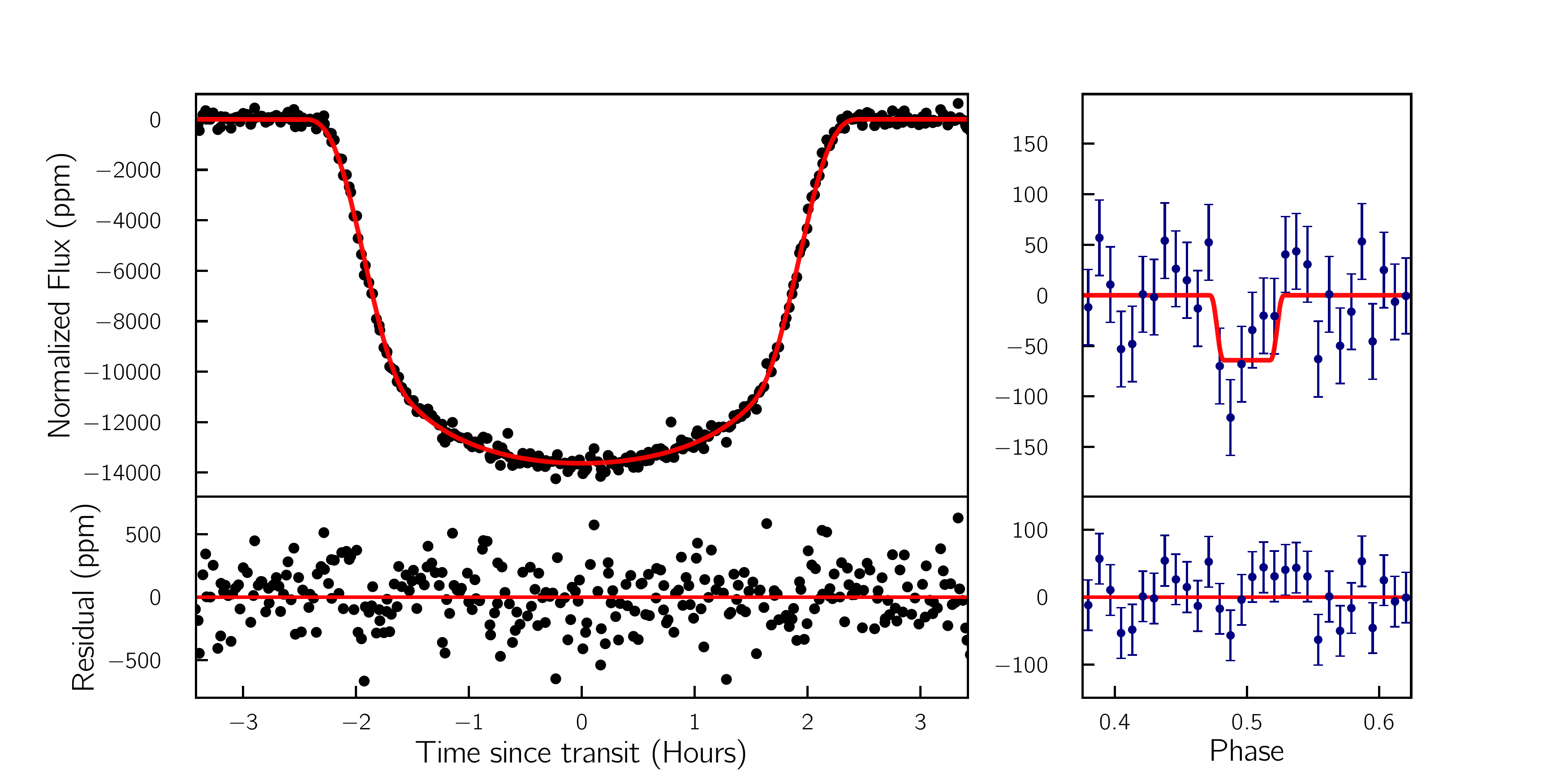}
\caption{\label{fig:hats_11b} Transit  fit (left) is shown for HATS-11b  using best fit parameters  which are reported in \autoref{table:secondaryeclp_param}. The secondary eclipse (right) is detected  at 62$\pm$12 ppm.}
\end{figure}
To our knowledge, K2-260b is the only planet up until now with a robust secondary eclipse detection among the planets discovered by \textit{K2}  \citep{johnson2018}. Since secondary eclipses characterize the geometric albedo as well as the temperature contrast just through observation of the depth of secondary eclipse alone, it is less prone to periodic or quasi-periodic noises. We therefore uniformly look for secondary eclipse signals for all planets reported in \autoref{table:best_candidates}.

For our secondary eclipse detection pipeline, we masked the range of data at transit as well as the phase of 0.5, where we estimate the secondary eclipse to occur. To the rest, we fitted a third degree polynomial with a length of 0.75 days and phase folded the data to build up the signal. The rest of the pipeline includes the same iterative outliers detection technique as was implemented for the phase curves. In this fashion, we detected secondary eclipse in \texttt{K2SFF} light curves of HATS-11b at 62$\pm$12 ppm (see \autoref{fig:hats_11b}). We use the depth to solve for the dayside temperature as well as the albedo for HATS-11b. The fit parameters are reported in \autoref{table:secondaryeclp_param}.

The detection of a secondary eclipse without the phase curve, as is the case for HATS-11b, raises an interesting question - why is there a secondary eclipse without a phase curve? HATS-11b has an easy-to-model stellar continuum that makes extracting a phase curve rather easy, although the signal may have been distorted during one of the many data processing steps. It also could be that the source of phase curve is predominantly thermal, and an efficient heat transportation between day and night-side significantly weakens the phase curve signal. Another explanation could be that the planetary atmosphere at the depth the phase curve probes is rotating at a pseudo-synchronous rate  \citep{adams2018} washing out the signal as we phase fold the light curve. Note using \texttt{EVEREST} light curve for HATS-11b, which has comparable  precision level as the \texttt{K2SFF} light curve, the secondary eclipse depth was detected at the level of 36$\pm$11 ppm, still a 3$\sigma $ detection. It still lacks a robust phase curve.

\section{Non-Detection}
\label{sec:nondetection}
For the most part, our formulated metric is expected to perform as  well as, if not better than, the previously used metrics such as a/R$_*<$10 in \citet{esteves2013}, and R$_p>$4R$_\oplus$, P$<$10 days and V$_{\mathrm{mag}}<$15 in \citet{angerhausen2015}. However, our precision approximation relation deviates from the actual observed value for fainter stars, which in the future could be improved by using the empirically obtained noise floor. Similarly our calculation of SNR ratio could underestimate the signal for small period planets due to potentially non-negligible contribution from tidal heating of the planets. 

\begin{figure}[ht]
\includegraphics[width=0.49\textwidth]{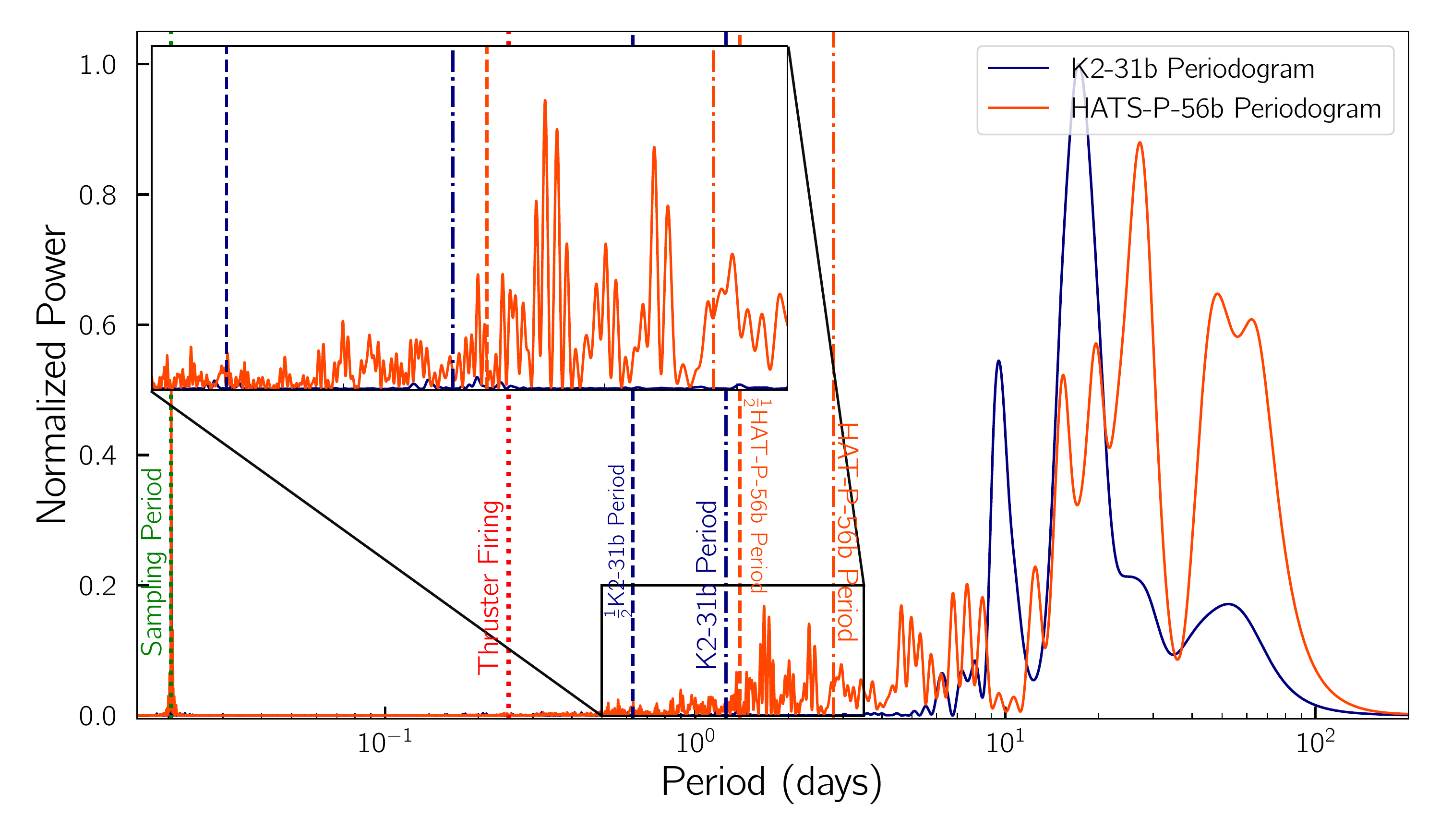}
\caption{\label{fig:comparativeperiodogram} Lomb-Scargle Periodogram of the time series of K2-31b  and HAT-P-56b after removal of outliers and transit points. Note the presence of significant power of HAT-P-56 near and around the planetary period while such signal is absent in case of K2-31b. The power at 6 hours (thruster firing) exhibits minimal power in both of the targets.}
\end{figure}

The presence of strong stellar activity can make the process of disentangling the phase curve difficult. The degree of difficulty particularly depends on the separation of the frequency and its harmonics of stellar modulation signals from the phase curve signal's frequencies. For instance, for targets for which phase curve signal was successfully extracted such as K2-31b, there is little overlap between power from stellar rotation and phase curve (see \autoref{fig:comparativeperiodogram}). On the other hand, for targets such as WASP-85Ab, K2-29b and K2-100b, the presence of prominent stellar activity dominates the frequency spectrum, thereby making the process of phase curve extraction difficult. For other targets such as WASP-118b and HATS-P-56b, the presence of stellar pulsation potentially induced by the planet similarly makes the disentangling process difficult  \citep{huang2015, mocnik2017}. A clear sign of stellar pulsation has been observed in similar spectral type but highly eccentric system -- HAT-P-2b \citep{dewit2017}. 

As for cases such as K2-137b, which has a large estimated ellipsoidal variation, no phase curve was detected because the mass was the upper limit reported in the discovery paper, leading to over-estimation of the signal \citep{smith2018}. Similarly, for targets such as K2-22b, a disintegrating planetary system \citep{ojeda2015}, the assumptions of our model does not hold. As for some of the targets such as WASP-47 system, we detect phase curve for the inner most planet at a significance level of 2$\sigma$, which we have not reported in this paper.

\section{Discussion}
\label{sec:Discussion}

\subsection{Model Performance}
The photometric mass  for the four hot Jupiters, while less precise than their RV counterparts, are consistent at 3$\sigma$ level with each other in all of the six cases. Hence, the model we use appear to be well-calibrated. In fact, the model has been tested for a wide range of masses -- planetary to stellar mass, although there are some cases of where discrepancy between the model and the data do exist  \citep{faigler2015b, eigmuller2018}. Such inconsistencies in the photometric mass in the case of hot Jupiters can arise due to the thermal component with a phase-shift resulting from super-rotation \citep{faigler2015a}. When we independently fit for ellipsoidal variation and Doppler boosting, the mass ranges derived from the Doppler effect were not as consistent with the RV mass as the one obtained from the ellipsoidal variation. For instance, the median of the distribution of ellipsoidal mass in all cases i.e. for Qatar-2b, WASP-104b, K2-31b, HATS-9b, and K2-107b was off compared to Doppler mass. While  such discrepancies could arise because of the fact that Doppler effects are often times too smaller compared to their ellipsoidal counterpart, and at the same time they are affected from the contribution of phase shifted thermal component or asymmetries in the reflective component. We refrained from using more complex models given the precision of data did not support such choices.  

\begin{figure}[ht!]
\centering
\includegraphics[width=0.49\textwidth]{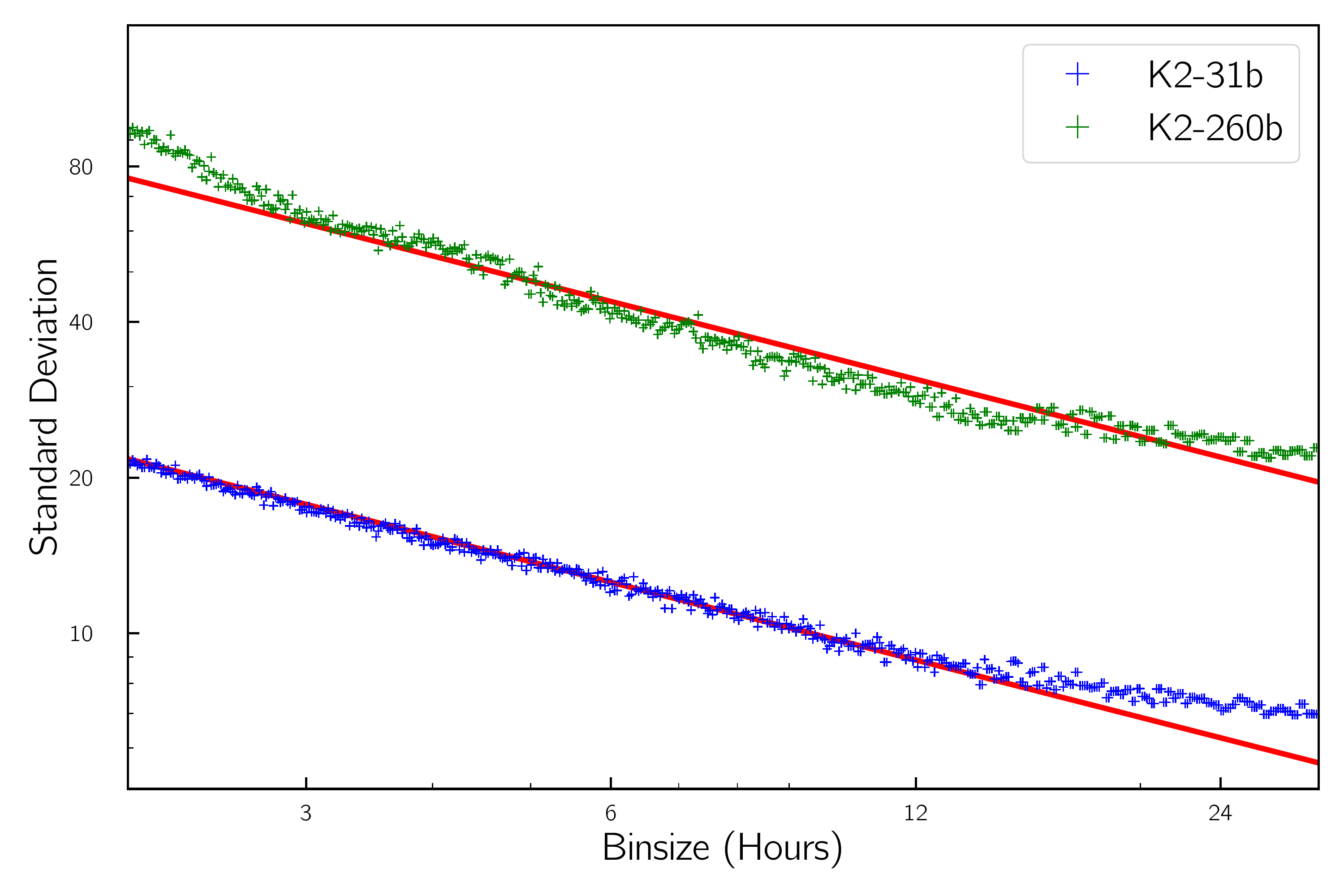}
\caption{\label{fig:allan_variance} Observed RMS vs the bin size in the residual obtained by fitting phase curve models in  K2-260b and K2-31b. Binning in K2-31b follows the expected power law of 0.5, whereas for K2-260b strongly deviates from it. The red lines are idealized cases drawn for both data sets.} 
\end{figure}

\subsection{Signal Fidelity}
\label{subsec:fidelity}
\textit{K2} has been used for used in studies related to astero-seismology  \citep{lund2016}, white dwarfs pulsation  \citep{hermes2017}, AGN variability \citep{aranzana2018}, and stellar rotation \citep{angus2016, esselstein2018}. These studies have shown that K2 pipeline can retain astrophysical variability signals if the period under consideration is less than 15 days \citep{vancleve2016}. As all of our targets have precisely known periods which fall well under 15 days, we can confidently extract the relevant signal. The major trouble in phase curve extraction instead is restricted by other dominant forms of accompanying stellar variation. 

There are additional tests we perform to test the fidelity of the signal. For instance, K2-260b is close to spin-orbit resonance. Despite reporting a prominent secondary eclipse, \citet{johnson2018} did not report the phase curve for the planet. We ran our pipeline, and fitted the phase folded light curve, which however yielded a mass inconsistent with reported RV value by more than 3$\sigma$. The residuals from the fit show the correlated noise which standout in RMS vs bin size plots compared to the residuals from the targets such as K2-31b for which the phase curve is reported (see \autoref{fig:allan_variance}). All of our phase curve targets have residuals that closely follow the expected the power law.

\subsection{Ultra-Short Period Planet}
With our discovery, K2-131b and K2-106b now join the group of other ultra-short planets such as Kepler-78 \citep{ojeda2013}, Kepler-10b \citep{esteves2013}, 55-Cnc-e \citep{demory2016} and K2-141b \citep{malavolta2018} with a detected secondary eclipse. Note all of these planets are rocky super-Earths with high densities and high geometric albedos possibly  due to the presence of refractory surfaces. We used the modified relation A$_g$=A$_B$ to estimate the geometric albedo for all of the super-Earth targets to allow geometric values greater than 0.66. Additionally, we suspect that there might be non-negligible additional source of heating such as tidal heating present in these systems which scales strongly with distance from the host star ($a$) and eccentricity ($e$) as follows: 
\begin{equation}
\begin{split}
    H = \frac{63}{4} \frac{(GM_*)^{3/2} M_* R_p^5}{Q_p} a^{-15/2}e^2\\
\end{split}
\end{equation}
where $H$ is the tidal heating rate, $M_*$ is the stellar mass, $R_p$ is the planetary radius, and $Q_p$ is the tidal dissipation parameter \citep{jackson2008}. The fact that most of these planets are multi-planetary system suggest mechanisms similar to Io as a Galilean moon of Jupiter may alse be acting on these planets \citep{peale1979, demory2016}. While tidal heating will increase the overall equilibrium temperature of these planets, thereby increasing the nightside contribution, the precision of \textit{K2} data does not allow us to explore such effects. Yet, asymmetries between the day and night-side can still occur due to mechanisms such as volcanism  \citep{gelman2011}, which would also contribute to the phase curve signals.

\subsection{Spectroscopic Follow-Up}
Phase curves can be used to infer the existence of atmospheres for the close-in hot planets through the detection of the  offset of the phase curve peaks \citep{shporer2015, demory2016, angelo2017}. Similarly, the geometric albedo can be linked to atmospheric processes such as clouds, which are known to play important roles in the  transmission spectrum \citep{kreidberg2014, sing2016}. Currently, there are only a few targets with reported geometric albedo which have been followed up with spectroscopic observation. However, this will drastically change as \textit{TESS} discovers a large sample of optimal targets. This could enable  screening out the best planetary candidates for the follow-up atmospheric studies using the geometric albedo as a guideline. 

\begin{deluxetable*}{lcccc}
\tablecaption{\label{table:secondaryeclp_param} Stellar and Planetary Parameters for K2-131b, K2-106b, and HATS-11b}
\tablehead{
\colhead{Parameter} &
\colhead{Unit} &
\colhead{K2-131b} &
\colhead{K2-106b} &
\colhead{HATS-11b} 
}
\startdata
\multicolumn{2}{l}{\textbf{Orbital Parameters}}\\
$M_*$ & $M_\odot$ & 0.84$\pm$0.03$^{a}$ & 0.92$\pm$0.03$^{b}$ & 1.000$\pm$0.060$^{c}$ \\
$R_*$  & $R_\odot$ & 0.81$\pm$0.03$^{a}$ & 0.95$\pm$0.05$^{b}$ & 1.444$\pm$0.057$^{c}$ \\
$T_{e\!f\!f}$  &K& 5200$\pm$100$^{a}$ & 5496$\pm$46$^{b}$ &  6060$\pm$150$^{c}$\\
$[$Fe/H$]$  & dex & -0.02$\pm$0.08$^{a}$ & 0.06$\pm$0.03$^{b}$& -0.390$\pm$0.060$^{c}$ \\
log $g_*$  & cgs & 4.62$\pm$0.10$^{a}$ & 4.42$\pm$0.05$^{b}$ & 4.118$\pm$0.026$^{c}$  \\
u & -& 0.6604$^{d}$ & 0.6294$^{d}$ & 0.5467$^{d}$\\
g & -& 0.4737$^{d}$& 0.4181$^{d}$ & 0.3199$^{d}$\\
\hline
\textbf{Pipeline}&-& \texttt{EVEREST} & \texttt{K2SFF} 
& \texttt{K2SFF}\\
\hline
\multicolumn{2}{l}{\textbf{Orbital Parameters}}\\
Period(Days) & Days &  0.3693038$\pm$0.0000091$^{a}$ & 0.571336$\pm$0.000020$^{b}$ & 3.6191613$\pm$ 0.0000099$^{c}$\\
$T_0$ -- 2450000 &  BJD & 7582.93620$^{+0.00031}_{-0.00033}$ & 6226.43381$^{+0.00040}_{-0.00039}$ &6574.96536$^{+0.00017}_{-0.00016}$\\
$R_p/R_*$ & - & 0.01968$^{+0.0016}_{-0.0006}$ & 0.01584$^{+0.00086}_{-0.00036}$&  0.10707 $\pm$0.00013 \\
$a/R_*$ &- & 2.61$^{+0.23}_{-0.58}$ & 2.73$^{+0.16}_{-0.47}$ &  7.006$^{+0.010}_{-0.015}$ \\
e & - & 0 (fixed) & 0 (fixed) & 0 (fixed) \\
$\omega$ & - & 90 (fixed) & 90 (fixed) & 0 (fixed)\\
b & - & 0.43$^{+0.32}_{-0.31}$ & 0.37$^{+0.31}_{-0.25}$ & 0.036$^{+0.039}_{-0.025}$\\
Inc & Deg & 80.5$^{+7.0}_{-12.4}$ & 82.3$^{+5.4}_{-9.7}$ &  89.70 $^{+ 0.21}_{- 0.32}$\\
u$_1$ & - & 0.511$^{+0.077}_{-0.054}$
 & 0.451$^{+0.072}_{-0.063}$ &  0.383$^{+0.007}_{-0.008}$\\
u$_2$ & -  & 0.141$^{+0.077}_{-0.056}$ & 0.228$^{+0.068}_{-0.068}$ &  0.2000$^{+0.0087}_{-0.0039}$ \\
\hline
\multicolumn{2}{l}{\textbf{Secondary Eclipse Fit Parameters}}\\
A$_{Re\!f + T\!h}$ & ppm & 12.7$\pm$4.2 & 11.7$\pm$2.5 & -\\
$\Delta$& ppm & 25.4$\pm$8.2 & 23.5$^{+4.9}_{-5.1}$ & 62$\pm$12\\
$A_g$&-& 0.27$^{+ 0.31}_{-0.27}$ & 0.62$^{+0.22}_{-0.34}$
& 0.249$^{+ 0.057}_{- 0.058}$\\
$T_{Day}$& K & 2300$^{+740}_{-425}$ & 2200$^{+630}_{-470}$ & 1704$^{+ 54}_{- 60}$
 \\
$T_{Eq}$ & K & 2010$^{+450}_{-270}$ & 1800$^{+470}_{- 375}$ & 1428$^{+ 44}_{- 49}$ 
\\
\enddata
\tablenotetext{a}{Adopted from \citet{dai2017a}}
\tablenotetext{b}{Adopted from \citet{sinukoff2017}}
\tablenotetext{c}{Adopted from \citet{rabus2016}}
\tablenotetext{d}{Interpolated from \citet{claret2011}}
\end{deluxetable*}

\subsection{Future Prospects}
With the launch of \textit{TESS} and the up-coming future missions like CHaracterising ExOPlanet Satellite  \citep[\textit{CHEOPS};][]{broeg2013} James Webb Space Telescope \citep[\textit{JWST};][]{beichman2014}, PLAnetary Transits and Oscillation of stars \citep[\textit{PLATO};][]{rauer2014}, and Atmospheric Remote-sensing Exoplanet Large-survey \citep[\textit{ARIEL};][]{tinetti2016}, there will be plentiful opportunities in the future for phase-curves studies. In fact, the phase curve of WASP-18b has already been reported  in  Sector 2 \textit{TESS} data \citep{shporer2018}, and  more will definitely be detected over the course of the mission. These studies will allow an unprecedented opportunity to learn about exoplanet atmospheres, while allowing us to refine our models with more precise data, and potentially disentangle the often degenerate reflective and the thermal components \citep{placek2016}.

\section{Conclusion}
We have significantly increased the number of phase curves discovered by \textit{K2} with four hot Jupiters' phase curves that yield photometric masses within 3$\sigma$ of the reported RV-based masses, and two additional short period super-Earths with 3$\sigma$  level  secondary eclipse detections along with corresponding phase curves. The availability of the precise light curves as well as the use of a more aggressive filtering procedure tested with signal injection facilitated in our venture. The consistency of the obtained mass, although for a small but non-negligible number of planets, raises the possibility of developing a tool for preliminary planetary signal validation for \textit{TESS} candidates. As we stand on the cusp of discovering many more planets, and will re-observe many of the hot Jupiters, an opportunity will be presented to refine our phase curve models, build a larger sample of planets with detected phase curves and open up novel lines of inquiry.  Such possibilities and others should strongly motivate the pursuit of the phase curves, as they will provide preliminary atmospheric characterization and mass estimation for many of the systems without investing any additional resources.

\vspace{0.5cm}

\noindent {\bf Acknowledgments:} This work includes data taken by \textit{K2}, and the final detrended light curves from \textit{K2SFF} as well as \texttt{EVEREST} pipeline. Authors would like to thank Dr. Andrew Vanderburg for discussion on \textit{K2SFF} pipeline products and kindly providing the detrended lightcurve of K2-141b. The authors would also like to thank Dr. Rodrigo Luger in regards to discussion on \texttt{EVEREST} pipeline. Authors would also like to thanks the anonymous referee for insightful comments. P. Niraula acknowledges the support of the Grayce B. Kerr Fellowship Fund. at MIT. S. Redfield and P. Niraula acknowledge support from the National Science Foundation through the Astronomy and Astrophysics Research Grant AST-1313268. D. Serindag acknowledges support from the European Research Council under the European Union's Horizon 2020 research and innovation program under grant agreement No. 694513. D. Serindag also acknowledges the Undergraduate Research Fellowship awarded by the NASA CT Space Grant Consortium in support of this research.

\software {\tt batman} \citep{kriedberg2015}, {\tt emcee} \citep{mackey2013}, {\tt gatspy} \citep{vanderplas2015}, {\tt lmfit} \citep{newville2016}, {\tt matplotlib} \citep{matplotlib}.

\appendix
The light curve of the six targets, where phase curve have been detected, are presented here. The cyan line represents the filtered light curve which was used to extract the phase curves. For K2-141b we use the light curve provided by Dr. Andrew Vanderburg which was used in their paper. For the rest of the eight cases, we use the light curve either from \texttt{EVEREST} or \texttt{K2SFF} as noted in their respective tables.

\begin{figure*}[ht!]
\centering
\includegraphics[width=0.90\textwidth]{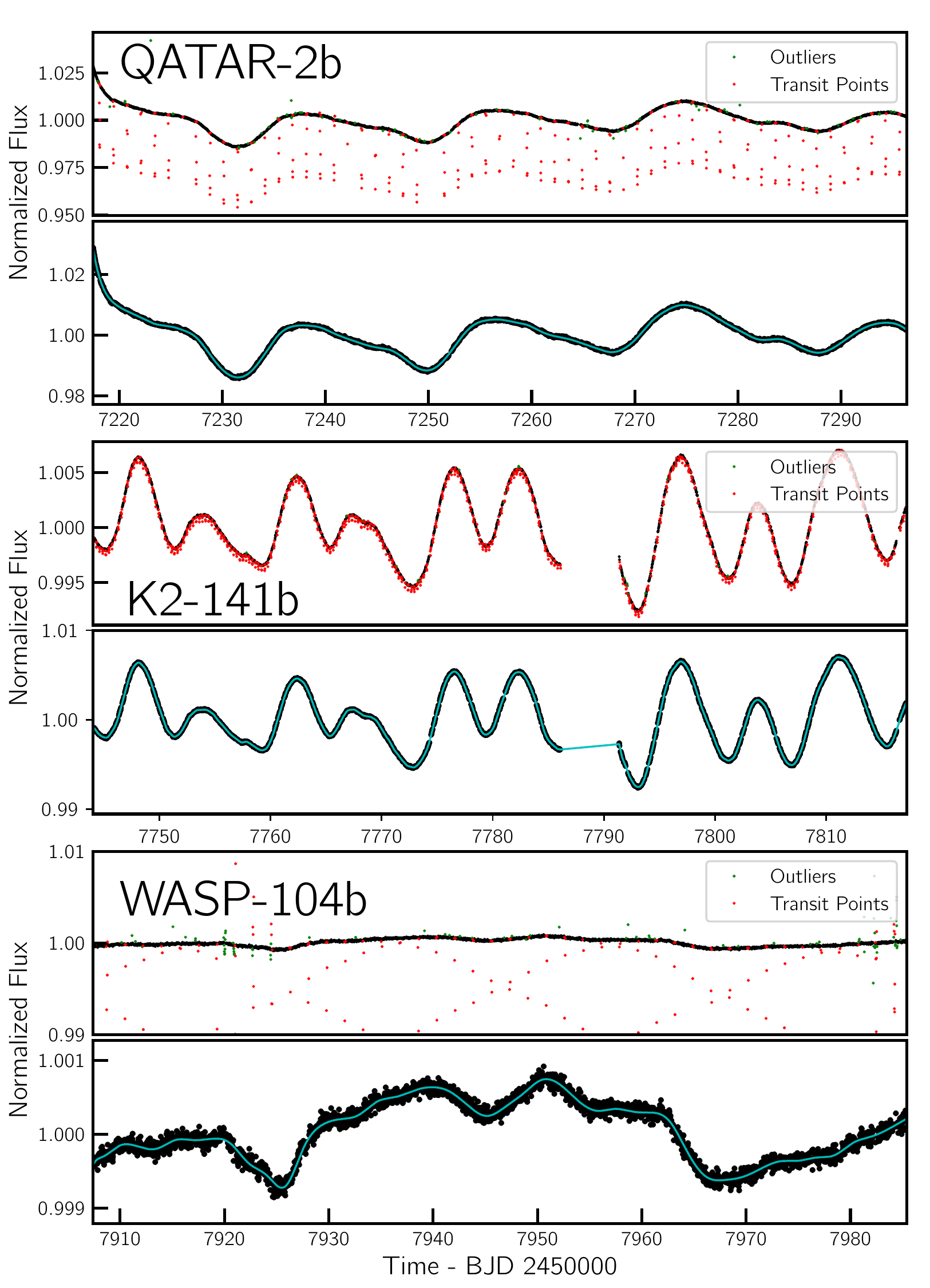}
\end{figure*}

\begin{figure*}[ht!]
\centering
\includegraphics[width=0.90\textwidth]{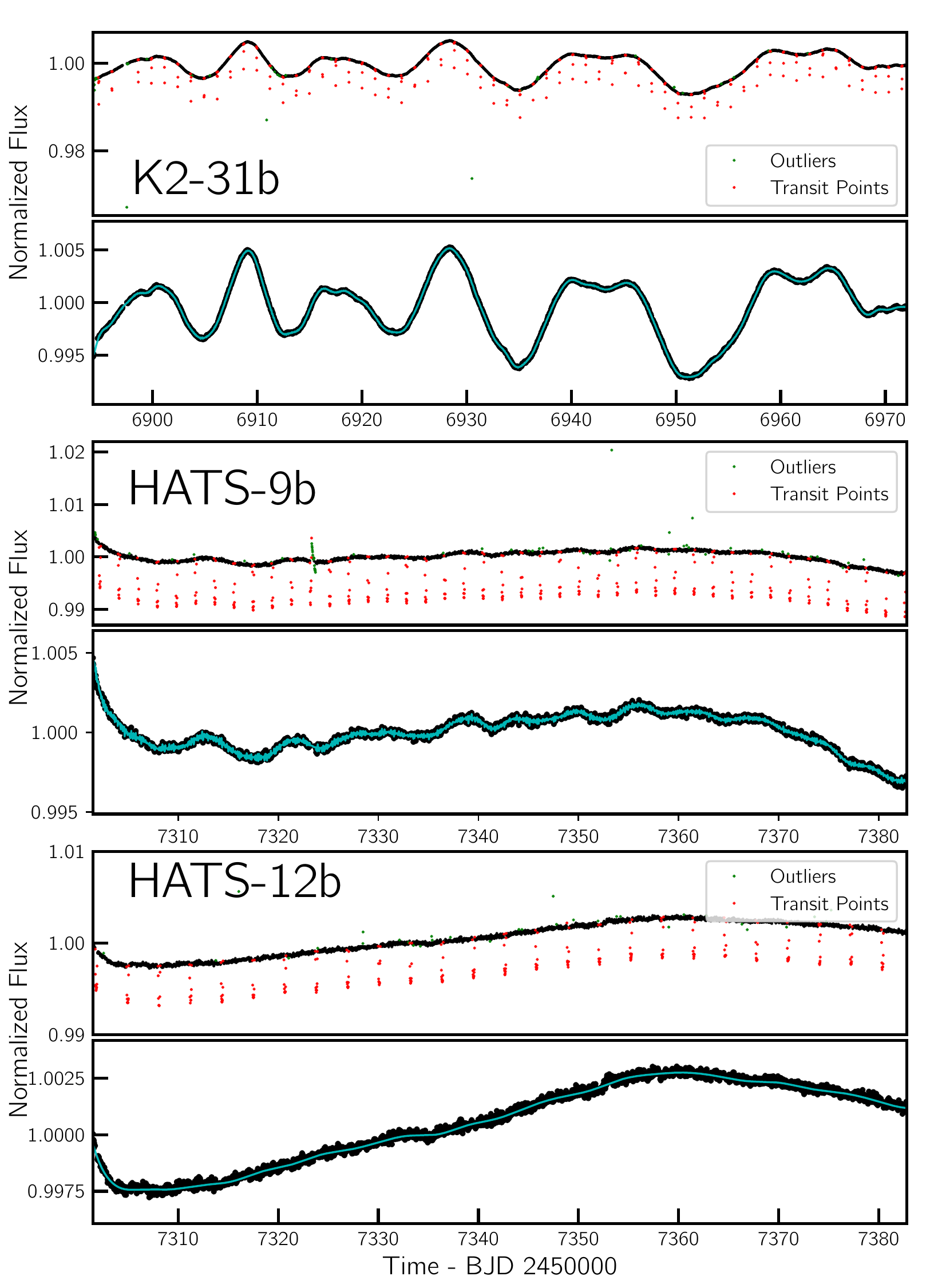}
\end{figure*}

\begin{figure*}[ht]
\centering
\includegraphics[width=0.90\textwidth]{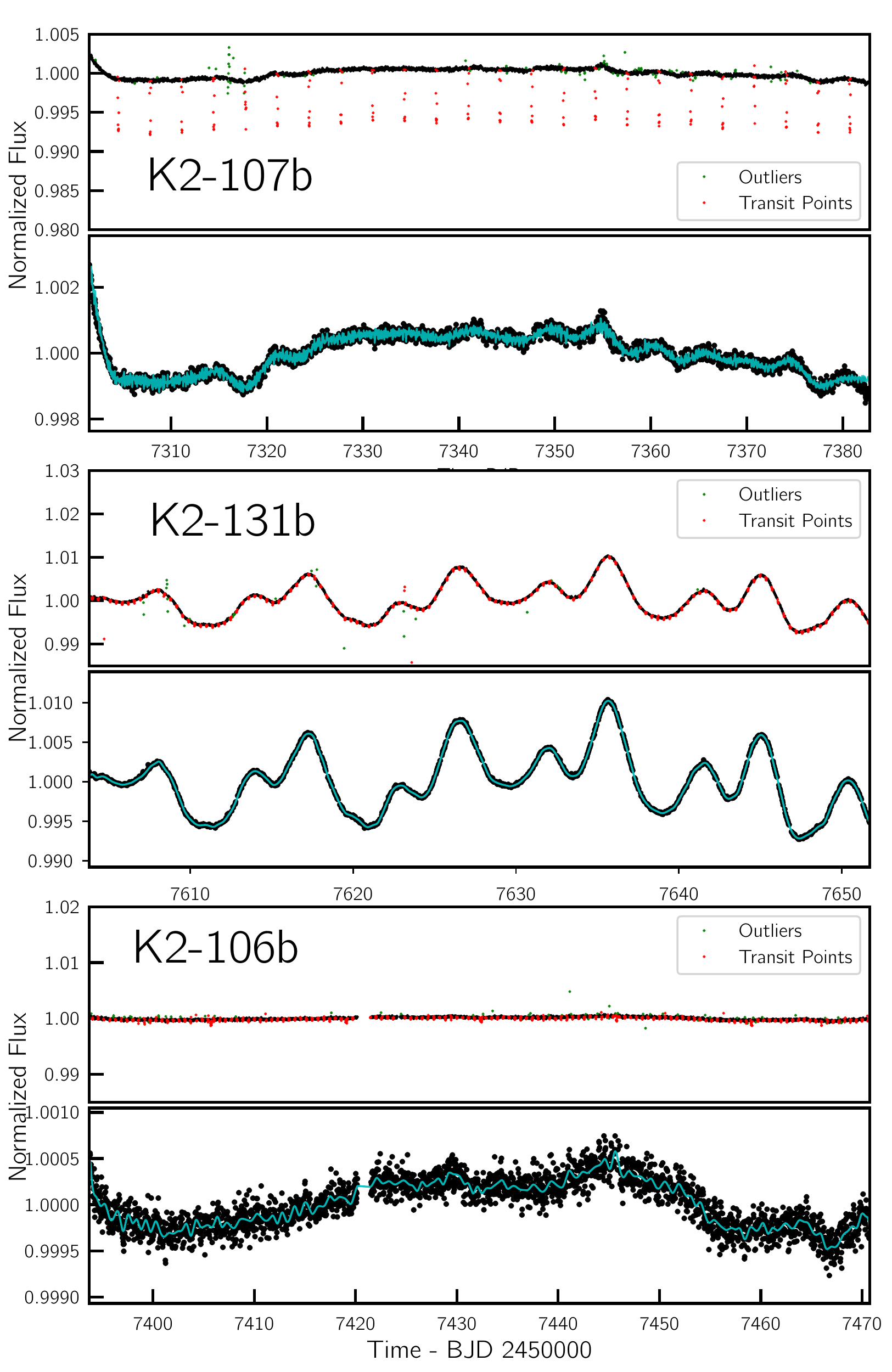}
\end{figure*}

\end{document}